\documentclass[11pt]{article}

\usepackage{color,graphicx}
\usepackage{young}
\usepackage[vcentermath]{youngtab}
\usepackage{amsmath,amssymb,graphicx,multirow}
\usepackage{hyperref}
\definecolor{darkred}{rgb}{0.65,0.15,0}
\hypersetup{pdfborder={0 0 0},colorlinks=true,urlcolor=darkred,citecolor=blue,linkcolor=darkred,linktocpage=true}

\usepackage{cite}
\usepackage{amsmath}
\usepackage{amsfonts}
\usepackage{amssymb}
\usepackage{graphicx}%
\usepackage{amsthm}
\usepackage{mathrsfs}
\usepackage[T1]{fontenc}
\usepackage{enumerate}
\usepackage{color}

\def\4diml{four-dimensional}

\def\-1{^{-1}}

\newcommand{\M}{\mathscr{M}}
\newcommand{\D}{\mathscr{D}}
\newcommand{\G}{\mathscr{G}}

\newcommand{\tG}{\widetilde{\mathscr{G}}}

\makeatletter

\@addtoreset{equation}{section}
\makeatother

\begin{document}

\thispagestyle{empty}

\vspace{5mm}

\begin{center}
{\Huge \bf Non-Abelian target space duals of \\[2mm] Thurston geometries}

\vspace{11mm}
\normalsize
~~~~~~{$ {\bf Ali ~Eghbali}$\footnote{Author to whom any correspondence should be addressed.},
$\bf{ Mahsa~ Feizi,~ Adel~ Rezaei}$-${\bf Aghdam}$}

\vspace{2mm}
~~~~~~{\small  Department of Physics, Faculty of Basic Sciences,
Azarbaijan Shahid Madani University, \\$~~~~~~~~~~~~~~~$ 53714-161, Tabriz, Iran}\\
\vspace{4mm}

\verb"Emails: eghbali978@gmail.com", \verb"mahsafeizi92818@gmail.com", \verb"rezaei-a@azaruniv.ac.ir"\\

\vspace{9mm}

\begin{tabular}{p{12cm}}
{\small
In this study, we proceed to investigate the Thurston geometries from the point of view of their Poisson-Lie (PL) T-dualizability.
First of all, we find all subalgebras of Killing
vectors that generate group of isometries acting freely and transitively on the three-dimensional
target manifolds, where the Thurston metrics are defined.
It is shown that three-dimensional Lie subalgebras are isomorphic to the Bianchi type algebras.
We take the isometry subgroup of the metric as the first subgroup of Drinfeld double.
In order to investigate the non-Abelian T-duality, the second
subgroup must be chosen to be Abelian. Accordingly, the non-Abelian target space duals of these geometries are found via PL T-duality approach in the
absence of $B$-field.
We also comment on the conformal invariance conditions of the T-dual $\sigma$-models under consideration.
}
\end{tabular}
\vspace{3mm}
\\
{\small {\it Keywords:} Non-linear $\sigma$-model, Non-Abelian T-duality, String duality, Thurston geometry}
\end{center}

\setcounter{page}{1}
\newpage
\tableofcontents

\section{\label{Sec.I} Introduction}

In mathematics, Thurston's geometrization conjecture states that each of certain
three-dimens-ional topological spaces has a unique geometric structure that can be associated with it.
It is an analogue of the uniformization theorem for two-dimensional surfaces, which states that every simply
connected Riemann surface can be given one of three geometries (Euclidean, spherical, or hyperbolic).
This theorem has proved to be a very powerful tool in two-dimensional physics, with applications in conformal field theories
and string theory. Unfortunately, there is no uniformization theorem in three dimensions, only a conjecture due to William Thurston
\cite{Thurston1980}. In fact, in three dimensions, it is not always possible to assign a single geometry to a whole topological space.
The conjecture proposed by Thurston  states that every closed three-manifold can be decomposed
in a canonical way into pieces that each have one of eight types of geometric structure \cite{Thurston1982}.
A three-dimensional model geometry \footnote{A model geometry is a simply connected smooth manifold
$\M$ together with a transitive action of a Lie group $G$ on $\M$ with compact stabilizers.
It is also called maximal if $G$ is maximal among groups acting smoothly and transitively on $\M$ with compact stabilizers.}$\M$
is relevant to the geometrization conjecture
if it is maximal and if there is at least one compact manifold with a geometric structure modelled on $\M$.
Thurston classified the eight model geometries satisfying these conditions,
each of which admits one, and only
one, of eight homogeneous geometries:  $H^3, S^3, E^3, E^1 \times S^2, E^1 \times H^2$, Sol, Nil and $\widetilde{SL}(2 , \mathbb{R})$.
The first three geometries are maximally symmetric, and hence isotropic. The remaining five
are anisotropic, and hence less symmetric, but all have at least a three-parameter group of
isometries. These geometries, which are called Euclidean Thurston geometries \cite{{Thurston1980},{Thurston1997},{Scott}},
their metrics are represented in Table \ref{table:1}.
It is worth noting that some Lorentzian metrics are obtained from Euclidean ones by a
Wick rotation \cite{Wick}. For instance, $M^3$ from $E^3$, de
Sitter from $S^3$, and anti-de Sitter ($AdS_3$) from $H^3$ or
$\widetilde{SL}(2 , \mathbb{R})$. However, since there is no distinguishable
direction over which to perform the rotation other inequivalent
metrics arise from the same Euclidean geometry. Following \cite{Gokhan},
here we represent Lorentzian Thurston geometries in seven model geometries called
Nil, $SL(2 , \mathbb{R})$, New Sol, Lorentz Sol, Third Sol, Lorentz-Heisenberg and $AdS_3$ (Table 2).
We also add the three-dimensional Lorentzian geometries $M^3$ and $AdS_2 \times S^1$ to the list given in \cite{Gokhan}.
A more complete list of Lorentzian left-invariant metrics has recently been given in \cite{chakkar1}.
The list contains three non-isometric classes of metrics for Nil geometry, for $SU(2)$ one class, for $\widetilde{PSL}(2, \mathbb{R})$
seven non-isometric classes, three and seven non-isometric classes for $\widetilde{E_0}(2)$ and Sol geometries, respectively.
Then, in Ref. \cite{chakkar2} it has been determined the isometry groups of all three-dimensional,
connected, simply connected and unimodular Lie groups endowed with a left-invariant Lorentzian metric.

On the other hand, it has been shown that the three-dimensional new massive gravity contains vacuum solutions with Thurston geometries
which have emerged as interesting three-manifolds with a given topology having canonical
decompositions into eight manifolds with high symmetry \cite{Flores1}, as well as arising as
the pictures in four-dimensional spatially homogeneous cosmological models \cite{{Taub},{Kantowski},{Fagundes}}
and have been used to construct five-dimensional black holes \cite{{Cadeau},{Hassaine1},{Hassaine2}} (see, also, \cite{Naderi.Rezaei}).
Lately, it has been shown that \cite{Flores2} four Lorentzian geometries
$M^3$, $AdS_3$, Nil and Sol that endow with left-invariant Lorentzian metrics on Lie groups,
can be solutions of three-dimensional general relativity non-minimally coupled to a scalar field and supplemented by electromagnetic matter.
It is noteworthy that early work on Thurston geometries as Euclidean gravitational
configurations came from string theory \cite{Gegenberg1}.
A stringy three-dimensional gravity theory was constructed as a way to
shed light on to the Thurston conjecture \cite{Gegenberg2}. An outstanding feature
of that work is that the gravity sector is given by
general relativity. Since most of the Thurston geometries
are not Einstein manifolds, then in previous analyses a
unifying dynamics required matter content.

The main purpose of this paper is to find the non-Abelian target space duals of Euclidean Thurston geometries via the PL T-duality approach.
We analyze their Lorentzian counterparts as well.
Klimcik and Severa \cite{{Klim1},{Klim2}} proposed PL T-duality
as a genuine generalization of Abelian \cite{Buscher} and non-Abelian dualities \cite{nonabelian}.
In contrast to its Abelian/non-Abelian descendants,
it does not require isomeric symmetries of the target space at all. Furthermore,
it deals with $\sigma$-models based on two Lie groups which form a Drinfeld
double \cite{Drinfeld} and the duality transformation exchanges their roles.
In recent years, we have witnessed further interest in PL T-duality, driven by $\sigma$-models based on Lie groups
\cite{{Alekseev1},{Tyurin},{Sfetsos1},{Lledo},{JR},{Majid},{N.Mohammedi2},{Hlavaty20002},{eghbali11},{EMR13},{Eghbali.2},{ERA1},{sakatani2}}.
In the present work, we shall investigate Thurston geometries from the point of view of their PL T-dualizability.
For that purpose, using the three-dimensional subalgebras\footnote{The numbering of the subalgebras follows
from the well-known classification of three-dimensional real Lie algebras \cite{bianchi}.
Non-isomorphic Lie algebras are written in 11 classes, $I, II, III, IV, V$, $VI_0, VI_a (a>0, a\neq1), VII_0, VII_a (a>0), VIII$
and $IX$, traditionally known as Bianchi algebras. Their commutation relations can be found in \cite{bianchi}.}
of the isometry Lie algebras that generate isometry subgroups acting {\it freely} and {\it transitively}
on the target space manifolds (metrics of the Thurston geometries)
we reconstruct the metrics as backgrounds of non-linear $\sigma$-models on the three-dimensional real Lie groups
of the Bianchi type.
For construction of dual backgrounds we use Drinfeld doubles obtained from the isometry subgroups of the metrics.
As mentioned earlier, our analysis focuses on geometries admitting free and transitive non-Abelian
isometry groups. We should be aware of the fact that many important target spaces in string theory
(notably coset spaces $G/H$) admit transitive action but non-free.
In fact, more general duality frameworks, such as those based on the double field theory or generalized cosets,
are capable of treating such cases \cite{F.Hassler}. However, we would like to point out that non-free manifolds can be dualized by using
the double field theory framework.

The procedure we will use in the present work was first applied for homogenous plane-parallel wave metric in \cite{Hlavaty1}
(see, also, \cite{{Hlavaty11},{hlavaty.ivo},{Petrasek.hlavaty.ivo},{hlavaty.ivo.Petrasek}}).
Then, it was applied for obtaining the non-Abelian duals of the $AdS_2 \times S^1$ and $AdS_3$ spaces \cite{ERA2}.
Recently, in Ref. \cite{ramirez} an $AdS_2$ solution to type IIA supergravity has been obtained by the non-Abelian T-duality
with respect to the free action of $SL(2,\mathbb{R})$ isometry group,  operating on
the $AdS_3 \times S^3 \times CY_2$ solution to type IIB. In this respect, the non-Abelian T-duality has been performed on the $AdS_3$.
In Ref.\cite{hlavaty.ivo}, all non-Abelian duals of the four-dimensional flat spacetime $M^4$ with respect to four-dimensional subgroups of
the Poincar\'{e} group have been classified. There, it has been obtained
14 different types of exactly solvable $\sigma$-models in the four-dimensional curved backgrounds.
In the present work, for the three-dimensional flat spacetime $M^3$ as one of the Lorentzian metrics
we find all corresponding non-Abelian duals. In this way, it is shown that the isometry Lie algebra of this spacetime encompasses
a large number of three-dimensional Lie subalgebras (all the Bianchi algebras except for $VI_a, VII_a$ and $IX$).

For convenience of the reader, we now provide an extended discussion of the
main steps of the approach used as well as an outline of the paper. The overall
task in order to achieve our goals mentioned above splits into five steps.
\\
\underline{step 1}: Calculating the Killing vectors of the metrics of Thurston geometries, we
obtain non-Abelian three-dimensional Lie subalgebras
of the isometry Lie algebras that generate isometry subgroups acting freely and transitively on three-dimensional target space manifolds where
the metrics of Thurston geometries are defined.
\\
\underline{step 2}: Since the dualizable metrics can be constructed by virtue of
Drinfeld double, the isometry subgroups of the metrics are taken as one of the subgroups of the Drinfeld double.
In order to satisfy the dualizability conditions the other subgroup is considered to be Abelian.
In other words, since we are dealing with non-Abelian T-duality,
the dual Lie algebra must be chosen to be Abelian.
\\
\underline{step 3}: In order to obtain the metrics of Thurston geometries by the construction of Drinfeld double
we need to find the transformation between the coordinates of group (isometry subgroup) and geometrical coordinates by choosing a convenient element of the group.
In this way, one needs to get the left-invariant vector fields on the group.
Accordingly, the metrics can be transformed into the group coordinates and finally
one can write the corresponding actions to the transformed metrics on the group which are nothing but the original $\sigma$-models.
\\
\underline{step 4}: Having the semi-Abelian Drinfeld doubles constructed by means of the non-Abelian isometry Lie subalgebras (with Abelian duals)
and then applying the PL T-duality transformation we obtain the dual $\sigma$-models whose backgrounds describe the non-Abelian target space duals of Thurston geometries.
\\
\underline{step 5}: Finally, we check the conformal invariance conditions of the mutually T-dual $\sigma$-models,
and show that some of them remain conformally invariant up to the one-loop order.
\begin{center}
	{ \scriptsize { \bf Table 1.}~{Euclidean Thurston geometries.}}\\
	\centering
	\begin{tabular}{| p{5em} | p{18em} |}
		\hline
		
		\scriptsize  {Geometry}  & \scriptsize {Metric} \\
		
		\hline
		
		\scriptsize $ E^3 $&\scriptsize $ ds^2=dx^2 +dy^2 +dz^2 $ \\
		
		\hline

         \scriptsize $ S^3 $&\scriptsize $ds^2=dx^2 + dy^2 + dz^2 +2 \cos x ~dy dz$ \\
		
\hline

 \scriptsize $ H^3 $&\scriptsize $ds^2=\frac{l^2}{z^2} \big(dx^2 + dy^2 +  dz^2\big)$ \\

\hline

 \scriptsize $ E^1 \times S^2 $&\scriptsize $ds^2={l^2} \big(dx^2 +dy^2 + \sin^2y ~dz^2\big)$ \\

\hline

\scriptsize $ E^1 \times H^2$&\scriptsize $ ds^2={l^2} \big(dx^2 +dy^2 + \cosh^2y dz^2\big)$ \\
		
\hline
\scriptsize Nil&\scriptsize $ ds^2=\frac{l^2}{4} \Big[dx^2 +dy^2 + (x dy-dz)^2\Big] $ \\
		
\hline

\scriptsize $ { \widetilde{SL}(2 , \mathbb{R})}$&\scriptsize $ds^2=\frac{1}{x^2} (dx^2 +2 dy^2)+ \frac{2}{x} dy dz+ dz^2$ \\

\hline

\scriptsize Sol &\scriptsize $ds^2={l^2}\Big[e^{-2z} dx^2 +e^{2z} dy^2 + dz^2\Big]$ \\

		\hline
	\end{tabular}
	\label{table:1}
\end{center}

\begin{center}
	{ \scriptsize {\bf Table 2.}~{Lorentzian Thurston geometries.}}\\
	\centering
	\begin{tabular}{| p{7em} | p{18em} |}
		\hline
		
		\scriptsize  {Geometry}  & \scriptsize {Metric} \\
		
     \hline
		
		\scriptsize $ M^3 $&\scriptsize $ ds^2=-dx^2 +dy^2 +dz^2 $ \\

		\hline
		
\scriptsize Nil &\scriptsize $ ds^2=\frac{l^2}{4} \Big[dx^2 +dy^2 - (x dy-dz)^2\Big]$ \\
		
		\hline
		
		\scriptsize $SL(2, \mathbb{R})$&\scriptsize $ ds^2=\frac{l^2}{4}  \Big[-(d\Psi  + \cos \Theta d\Phi)^2 + d\Theta^2 +\sin^2\Theta~ d\Phi^2\Big]$ \\
		
		\hline
         \scriptsize New Sol&\scriptsize $ds^2={l^2}\Big[e^{-2z} dx^2 +e^{2z} dy^2 - dz^2\Big]$ \\
		
\hline

 \scriptsize Lorentz Sol&\scriptsize $ds^2={l^2}\big(2e^{-z} dx dz +e^{2z} dy^2\big)$ \\

\hline
         \scriptsize Third Sol&\scriptsize  $ds^2={l^2}\big(-e^{2z} dy^2- 2 dx dy + dz^2\big)$ \\
		
\hline

\scriptsize Lorentz-Heisenberg&\scriptsize  $ds^2=\frac{l^2}{4} \Big[-dx^2 + dy^2+(x dy -dz)^2\Big]$ \\

\hline

\scriptsize $ AdS_2 \times S^1 $&\scriptsize $ds^2=l^2 dx^2 + \frac{l^2}{z^2} \big(-dy^2 + dz^2\big)$ \\

\hline
\scriptsize $AdS_3$&\scriptsize $ds^2=\frac{l^2}{z^2} \big(- dx^2+ dy^2 +dz^2\big)$ \\

		\hline
	\end{tabular}
	\label{table:2}
\end{center}

We have organized our manuscript as follows:
After Introduction section, section \ref{Sec.II} reviews the construction of PL T-dual $\sigma$-models on Lie groups, where necessary formulas are summarized.
The discussion of the non-Abelian T-duality via PL T-duality approach is also given at the end of this section.
In section \ref{Sec.III}, we calculate the Killing vectors of Thurston geometries as the generators of isometries, and then classify
the non-Abelian isometry subgroups acting freely and transitively on these geometries; the results are summarized in Tables 3-6.
Section \ref{Sec.IV} contains the original results of the work:
using the dualization procedure described in section \ref{Sec.II}, we construct the
non-Abelian target space duals of Thurston geometries. The results including
the constant matrix $E_0(e)$, the transformation between the coordinates of Thurston metrics and group ones, together with the
metrics and $B$-fields corresponding to both original and dual backgrounds are summarized in Tables 7 and 8.
Detailed discussion of particular example, the non-Abelian T-dualization of the Lorentz Sol geometry, is also given in section \ref{Sec.IV}.
We start section \ref{Sec.V} by introducing the vanishing of the beta-function equations up to the one-loop order, and then
study the conformality of the T-dual $\sigma$-models up to the one-loop order.
We conclude with a final discussion of the results with remarks and perspectives.
The isomorphism transformation between six-dimensional Drinfeld doubles with
the isometry Lie algebras of the $E^3$, $H^3$ and $M^3$ geometries is left to Appendix A.


\section{\label{Sec.II} Non-Abelian T-duality via PL T-duality approach}

We begin with a brief summary of the construction of PL T-dual $\sigma$-models on Lie groups \cite{Klim1,Klim2}.
In what follows, first we consider a two-dimensional non-linear $\sigma$-model on a manifold $\M$ as target space with some scalar fields $x^\mu$, where
$x^\mu,~\mu =1,...,~$dim(G) parametrize an element $g$ of a Lie group $G$. We also introduce representation matrices $\{T_a\}$ as basis of the Lie algebra $\G$ of $G$, with $a = 1,...,$ dim(G) and the components of the right-invariant Maurer-Cartan one-forms
$R_{\pm}^a = (\partial_{_\pm} g~g^{-1})^a = \partial_{_\pm} x^{\mu}~ R_{\mu}^{~a}$. The inverse of $R_{\mu}^{~a}$
will be denoted by $R_{a}^{~\mu}$.
The light-cone coordinates on the world-sheet are $\sigma^{\pm} =(\tau \pm \sigma)/2$ together with
$\partial_{_\pm}=\partial_{\tau} \pm \partial_{\sigma}$.
The corresponding action in the light-cone coordinates has the form
\begin{eqnarray}\label{sec2.2}
S = \frac{1}{2}\int \!d\sigma^+  d\sigma^- ~ ({G}_{\mu\nu}+ {B}_{\mu\nu})  \partial_{_+} x^\mu \partial_{_-} x^\nu,
\end{eqnarray}
where the combination of metric ${G}_{\mu\nu}$ and  Kalb-Ramond field ${B}_{\mu\nu}$ ($B$-field)
together defines the tensor field ${\cal E}_{\mu \nu}$.
We also define the corresponding line element and $B$-field, in the coordinate basis, as
$ds^2 = G_{\mu\nu} dx^\mu dx^\nu$ and $B = {1}/{2} ~B_{\mu \nu} ~ dx^\mu \wedge dx^\nu$.

Now, suppose that the Lie group $G$ acts from right on $\M$  freely. If the Noether's current one-forms corresponding to the right action of
$G$ on $\M$ are not closed and satisfy the Maurer-Cartan equation with the structure constants of the Lie group $\tilde G$
(the dual Lie group to $G$ with the same dimension $G$) on the
extremal surfaces, we say that the $\sigma$-model \eqref{sec2.2} has the PL symmetry
with respect to the $\tilde G$. It is
the condition of dualizability of $\sigma$-models on the level of the Lagrangian which is given by \cite{{Klim1},{Klim2}}
\begin{eqnarray}\label{sec2.3}
{\cal L}_{_{V_{a}}}{\cal E}_{\mu \nu}={{\tilde f}^{bc}}_{~a}~ {\cal E}_{_{\mu \sigma}}~ {V_{c}}^{{~\sigma}}~{V_{b}}^{{~\lambda}}~
{\cal E}_{{\lambda\nu}},
\end{eqnarray}
where ${\cal L}_{_{V_{_a}}}$ stands for the Lie derivative corresponding to the left-invariant vector
fields ${V_{a}}= {V_{a}}^{\mu} {{{\partial}/{\partial{x^{\mu}}}}}$ constructing on the Lie group $G$, and
${{\tilde f}^{bc}}_{~~a}$ are the structure constants of the dual Lie algebra $\tG$ of $\tilde G$.
The Lie algebras $\G$ and $\tG$ then define the Drinfeld double that enables to construct tensor ${\cal E}_{\mu \nu}$ satisfying \eqref{sec2.3}.

Since the PL duality is based on the concepts of the Drinfeld double, it is necessary to define the Drinfeld double group $D$.
A Drinfeld double \cite{Drinfeld} is simply a Lie group $D$ whose Lie algebra $\D$
admits a decomposition $\D =\G \oplus {\tilde \G}$ into a pair of subalgebras maximally isotropic
with respect to a non-degenerate ad-invariant symmetric bilinear form $<.~,~.>$.
The dimension of subalgebras have to be equal. We furthermore consider $G$ and $\tilde G$ as a pair of maximally isotropic subgroups corresponding
to the subalgebras $\G$ and $\tilde \G$,
and choose a basis in each of the subalgebras as
$T_{{_a}} \in \G$  and ${\tilde T}^{a} \in {\tilde \G}, a = 1, ...,$ dim(G), such that
\begin{eqnarray}\label{sec2.4}
<T_{{_a}} ,  T_{{_b}}> = <{\tilde T}^{{^a}} ,  {\tilde T}^{{^b}}> =0,~~~~~~~~<T_{{_a}} , {\tilde T}^{{^b}}>  = {\delta}_{_{a}}^{{~b}}.
\end{eqnarray}
The basis of the two subalgebras satisfy the commutation relations
\begin{eqnarray}\label{sec2.5}
[T_a , T_b] = {f^{c}}_{ab} ~T_c,~~~~~[{\tilde T}^{a} , {\tilde T}^{b}] ={{\tilde f}^{ab}}_{~~c} ~{\tilde T}^{c},~~~~
[T_a , {\tilde T}^{b}] = {{\tilde f}^{bc}}_{~~a} {T}_c + {f^{b}}_{ca} ~{\tilde T}^{c},
\end{eqnarray}
where ${f^{c}}_{ab}$ and $\tilde f^{ab}_{~~c}$ are structure constants of $\G$ and $\tilde \G$, respectively\footnote{
The Jacobi identity on $\D$ relates the structure constants of the two Lie
algebras as
\begin{eqnarray}\label{sec2.6}
{f^a}_{bc}{\tilde{f}^{de}}_{\; \; \; \; a}=
{f^d}_{ac}{\tilde{f}^{ae}}_{\; \; \; \;  b} +
{f^e}_{ba}{\tilde{f}^{da}}_{\; \; \; \;  c}+
{f^d}_{ba}{\tilde{f}^{ae}}_{\; \; \; \; c}+
{f^e}_{ac}{\tilde{f}^{da}}_{\; \; \; \; b}.
\end{eqnarray}
}.
Note that the Lie algebra structure defined by relation \eqref{sec2.5} is called Drinfeld double $\D$.

In the following, we shall consider T-dual $\sigma$-models on a Drinfeld
double $D$. To this end, we must assume that $G$ acts transitively and freely on the manifold $\M$;
then the target can be identified with the Lie group $G$, $\M \approx G$.
Under the condition \eqref{sec2.3} the field equations of the model \eqref{sec2.2} can be
rewritten as equation for the mapping $l(\sigma^+, \sigma^-)$ from the world-sheet into the double $D$ \cite{{Klim1},{Klim2}}
\begin{eqnarray}\label{sec2.7}
<\partial_{\pm} l~l^{-1} , \varepsilon^{\mp}>=0,
\end{eqnarray}
where $l \in D$, and subspaces $\varepsilon^{+}$=span$\{T_{a} + {E_{_0}}_{ab} {\tilde T}^{{b}} \}$ and
$\varepsilon^{-} $=span$\{T_{a} - {E_{0}}_{{ba}} {\tilde T}^{{b}} \}$ are orthogonal with respect to
the $<.~ , ~.>$ and span the whole Lie algebra ${\D}$. Note that ${E_{_0}}_{ab}$ is an invertible
constant matrix at the vicinity of the unit element of G, $g=e$,
which is called the $\sigma$-model constant matrix.
One can write $l = g {\tilde h}; g \in G, {\tilde h} \in {\tilde G}$ (such
decomposition of group elements exists at least at the vicinity of the unit element); moreover, one may write
$l = {\tilde g} h$ where  $h \in G, {\tilde g} \in {\tilde G}.$
In this case, the original $\sigma$-model having target space in the Lie group $G$ is given by the following action
\begin{eqnarray}\label{sec2.8}
S=\frac{1}{2}\int\!d\sigma^{+} d\sigma^{-}  {{R_{_+}}}^{\hspace{-1.5mm} a}\;{{R_{_-}}}^{\hspace{-1.5mm} b} {E_{_{ab}}}(g),
\end{eqnarray}
in which ${E_{_{ab}}}(g)$ is a certain bilinear form on the $G$ and is defined by
\begin{eqnarray}\label{sec2.9}
 {E_{_{ab}}}(g) = \big(E_{_0}^{-1} + \Pi (g)\big)^{-1},~~~~~~\Pi^{ab}(g)=b^{ac}(g) ~ (a^{-1})_c^{~b}(g),
\end{eqnarray}
where $\Pi(g)$ is the Poisson structure on the $G$, and the submatrices $a(g)$ and $b(g)$ are the adjoint representations of the $G$
on $D$ in the basis $(T_{{a}}, {\tilde T}^{a})$ which are defined in the following way
\begin{eqnarray}\label{sec2.10}
g^{-1} T_{{a}}~ g &=&a_{_{a}}^{^{~c}}(g) ~ T_{{c}},\nonumber\\
g^{-1} {\tilde T}^{{a}} g &=& b^{^{ac}}(g)~ T_{{c}}+(a^{-1})_c^{~a}(g)~{\tilde T}^{{c}}.
\end{eqnarray}

It is possible to define an equivalent but dual $\sigma$-model by the exchange of $G \leftrightarrow {\tilde G}$,
$\G \leftrightarrow {\tG}$, $E_{_0} \leftrightarrow {\tilde E}_{_0} =E_{_0}^{-1}$ and $\Pi(g) \leftrightarrow {\tilde \Pi}({\tilde g})$.
Accordingly, we consider another $\sigma$-model on the dual Lie group ${\tilde {G}}$, whose dimension is
equal to that of $G$, with the coordinates
${\tilde x}^{\mu},~ \mu = 1,...,$ dim($\tilde G$)
which parameterize an element $\tilde g$ of the ${\tilde {G}}$.
The corresponding $\sigma$-model has the form
\begin{eqnarray}\label{sec2.11}
\tilde S = \frac{1}{2} \int d\sigma^{+}  d\sigma^{-}~ {\tilde R}_{+_{a}}{\tilde R}_{-_{b}} {{{\tilde E}}^{{ab}}}(\tilde g).
\end{eqnarray}
where ${\tilde R}_{{\pm}_a}$'s defined by
$( \partial_{_\pm} \tilde g ~\tilde g^{-1})_a=\partial_{\pm} {\tilde x}^{\mu} {\tilde R}_{\mu a}$
are the components of the right-invariant Maurer-Cartan one-forms on the $\tilde G$.
Also, the background of the dual theory is related to that of the original one by
\begin{eqnarray}\label{sec2.12}
{{\tilde E}}(\tilde g) = \big(E_{0}+ {\tilde \Pi}(\tilde g)\big)^{-1},
\end{eqnarray}
where ${\tilde \Pi}(\tilde g)$ is defined as in \eqref{sec2.9} by replacing untilded quantities with tilded ones.
Note that the actions \eqref{sec2.8} and \eqref{sec2.11} correspond to PL T-dual $\sigma$-modes\cite{Klim1,Klim2}.

Before closing this section, let us discuss the non-Abelian T-duality case of T-dual $\sigma$-models \eqref{sec2.8} and \eqref{sec2.11}.
In this case we will be dealing with a semi-Abelian double for which we consider ${f^a}_{bc} \neq 0$ and $\tilde f^{ab}_{~~~c}=0$.
Then, it follows from \eqref{sec2.5} and \eqref{sec2.10} that $ b(g)=0 $, then, $\Pi(g)=0 $; consequently, $ E=E_0$, and thus the action
\eqref{sec2.8} becomes\footnote{When the matrix ${E_{_0}}_{ab}$ is replaced with
a non-degenerate metric on Lie algebra $\G$ of $G$, the model becomes a principal chiral one.}
\begin{eqnarray}\label{sec2.13}
S=\frac{1}{2}\int\!d\sigma^{+} d\sigma^{-}  {{R_{_+}}}^{\hspace{-1.5mm} a}\;{{R_{_-}}}^{\hspace{-1.5mm} b} {E_{_0}}_{ab}.
\end{eqnarray}
By identifying the above action with the $\sigma$-model action of the form \eqref{sec2.2},
one can obtain the relationship between ${\cal E}_{\mu \nu}$ and ${E_{_0}}_{ab}$. It is then given by the formula
\begin{eqnarray}\label{sec2.14}
{\cal E}_{\mu \nu} = R_{\mu}^{~a} R_{\nu}^{~b} {E_{_0}}_{ab}.
\end{eqnarray}
We should note that in this case of the non-Abelian T-duality, the action \eqref{sec2.11} remains unchanged.
In addition, let us note that if ${E_{_0}}_{ab}$ is chosen to be symmetric, then one concludes that the $B$-field  vanishes.
In general, the matrix ${E_{_0}}_{ab}$ can have an anti-symmetric part which in that case the $B$-field would be non-vanishing.
In the following, we will construct Thurston geometries from action \eqref{sec2.13} by a suitable choice of the matrix $E_{_0}$. Then,
by applying the PL T-duality transformations together with action \eqref{sec2.11} we obtain the corresponding non-Abelian T-dual backgrounds.

\vspace{-2mm}

\section{\label{Sec.III} The free and transitive action of non-Abelian isometry subgroups on Thurston geometries}

In this section, we calculate the Killing vectors of Thurston geometries as the generators of isometries, and then classify
the non-Abelian isometry subgroups acting freely and transitively on these geometries.
As mentioned in the Introduction, the isometry subgroups can be used
for construction of the non-Abelian T-dual backgrounds. Sufficient condition for that is
that the metric has a non-Abelian isometry subgroup whose dimension is equal to
the dimension of the manifold and its action on the manifold is free and also transitive.
Of course, it goes without saying that the relation between the model geometries of Table 1 and the Bianchi type has already been studied in \cite{hervik}.

In order to construct dualizable backgrounds for Thurston geometries
we need to first investigate whether the action of the non-Abelian isometry subgroups on these geometries is free and transitive.
Before proceeding further, let us have some review on the definition of free and transitive actions \cite{Nakahara}.
The free action of a Lie group $G$ on a manifold $\M$ is defined by
$ g\circ x^\mu=x^\mu \Rightarrow g=e$,
where $g\in  G$ and  $x^\mu$ denotes any point of ${\M}$.
Consider $g=e^{\alpha^a T_a}$ where $T_a$'s are the bases of Lie algebra of $G$,
and $\alpha^a$'s being some constant parameters. Then, the infinitesimal form of this action leads us to $ \xi \circ x^\mu=0 $ in which $ \xi=\alpha^a T_a $.
In order to have the free action one must expand the bases $T_a$ in terms of the Killing vectors of the manifold metric
as $T_a = A_a^{~\mu} \partial_{_\mu}$.
Then, by applying $ \xi \circ x^\mu=0 $ one concludes that $ \alpha^a=0 $, that is, $g=e$.
In order to have the transitive action, for a non-trivial element $ g $ of $G$
we must have $g \circ x^\mu={x'}^\mu $ where $x^\mu$ and ${x'}^\mu$ are two points of $\M$.
The Killing vectors of $\M$ must form a basis of tangent space in every point of $\M$. It means that in every point of $\M$
there is an invertible matrix $A_a^{~\mu} (x^\nu)$ that solves the equation
\begin{eqnarray}
T_a=A_a^{~\mu} \partial_\mu.\label{3.1}
\end{eqnarray}
Note that linear independence of the bases $T_a$ requires that matrix $A_a^{~\mu}$ be invertible.
In addition, we must pay attention to the fact that if the action of ${G}$ on the $\M$ is free and also transitive,
then the dimensions of both $G$ and $\M$ must be equal.

As was mentioned earlier, the Killing vectors of Thurston geometries give us the basic data for dualization.
First of all, from solving Killing equations, $ \mathcal{L}_{_{k_{a}}}G_{{\mu \nu}}=0$, we calculate the Killing vectors of Thurston geometries along with the commutation relations of the isometry Lie algebras.
The results for both Euclidean and Lorentzian Thurston geometries are summarized in Tables 3 and 4, respectively.
As can be seen of Tables 3 and 4, all metrics of Thurston geometries, except for Sol, New Sol and Third Sol,
possess isometry group whose dimension is greater than the dimension of the manifold. Accordingly, we may have several duals for these metrics.
However, one can obtain only one dual pair for the metrics Sol, New Sol and Third Sol, because their isometry Lie algebra is
only the Bianchi type algebra $VI_0$.
The most interesting case is the $M^3$ geometry whose isometry Lie algebra encompasses a large number of three-dimensional Lie subalgebras (all the Bianchi algebras except for $VI_a, VII_a$ and $IX$).
It is worth stressing that, the non-Abelian isometry Lie subalgebras of Killing vectors that generate subgroup of isometries which acts freely
and transitively on the manifolds defined by the $AdS_2 \times S^1$ and $AdS_3$ geometries, have been given in \cite{ERA2}. There,
it has been obtained all non-Abelian dual backgrounds for these two families of $AdS$ spaces.
Accordingly, from now on these two  geometries will be excluded from our investigations.
Note that one can show that the isometry Lie algebras of the $E^3$, $H^3$ and $M^3$ geometries are isomorphic to the Lie algebras of Drinfeld doubles
$ {{\cal D} {\cal D}}_5:(IX|I)$, $ {{\cal D} {\cal D}}_1:(IX|V|b), b>0$ and $ {{\cal D} {\cal D}}_7:{(VII_0|IV|b)}_{_{b=1}}$, respectively \cite{Snobl.Hlavaty}.
The isomorphism transformations between these Drinfeld doubles with
the isometry Lie algebras of the aforementioned geometries are given in Appendix A.


\begin{center}
		\scriptsize {{{\bf Table 3.}~ Killing vectors and isometry Lie algebras of Euclidean Thurston geometries.}}
		{\scriptsize
			\renewcommand{\arraystretch}{1.5}{
\begin{tabular}{| l|l|l| } \hline \hline
Geometry  & Killing vectors & Lie algebra of  \\  &   &  isometry group\\ \hline
          &       $k_{1} = y \partial_{x}-x \partial_{y} $  &  $ {\bf {{\cal D} {\cal D}}_5:~(IX|I)} $ \\
     &  $k_{2}=  z \partial_{x}-x \partial_{z},$ &  $[k_{1} , k_{4}]=-k_{2},~[k_{1} , k_{5}]= -k_{3},$\\
 $E^3$ &      $k_{3}=\partial_{x},$ & $[k_{2} , k_{3}]=k_{6},~ ~[k_{2} , k_{4}]=k_{1},$ \\
     &      $k_{4}=  z \partial_{y}-y \partial_{z},$ & $[k_{2} , k_{6}]=-k_{3},~ [k_{4} , k_{5}]=k_{6},$ \\
     &    $k_{5}= \partial_{y},$  & $ [k_{4} , k_{6}]= -k_{5},~[k_{1} , k_{3}]= k_{5}$ \\
         &    $k_{6}=  \partial_{z}$  & $[k_{1} , k_{2}]=k_{4}$ \\\hline

    &    $k_{1}= \sin z ~\partial_{x}- {\cos z}~{\csc x} ~ \partial_{y} +{\cot x~\cos z} ~ \partial_{z}, $  &  ${\bf so(3) \oplus so(3)}: $ \\
    &    $k_{2}= \cos z ~\partial_{x}+ {\sin z}~{\csc x} ~\partial_{y}-{\cot x~ \sin z} ~ \partial_{z}, $  & $[k_{1} , k_{2}]=k_{6}, ~[k_{1} , k_{6}]= -k_2,$ \\
$S^3$ & $k_{3}= \sin y ~\partial_{x}+{\cot x}~{\cos y }~\partial_{y} -{\csc x~ \cos y}~\partial_{z}, $  &  $[k_{2} , k_{6}]=k_{1}, ~[k_{3} , k_{4}]= k_{5},$\\
    &    $k_{4}= \cos y ~\partial_{x}- {\cot x}~{\sin y }~ \partial_{y}+{\csc x~ \sin y} ~\partial_{z}, $ & $[k_{3} , k_{5}]=-k_{4}, ~[k_{4} , k_{5}]= k_{3}$ \\
    &    $k_{5}=  \partial_{y},~~~~k_{6}= \partial_{z}$  &\\\hline

     &       $k_{1}= \frac{1}{2} (x^2-y^2-z^2)\partial_{x} +xy \partial_{y} +xz \partial_{z},$  & ${\bf {{\cal D} {\cal D}}_1:~(IX|V|b),~~b>0 }$   \\
$H^3$ &  $k_{2}= -\frac{1}{2} (x^2-y^2+z^2)\partial_{y} +xy \partial_{x} +yz \partial_{z},$ &  $[k_{1} , k_{3}]=-k_{1}, ~[k_{1} , k_{4}]=- k_{2},$\\
  &      $k_{3}= x \partial_{x}+y \partial_{y}+z \partial_{z},$ &   $[k_{1} , k_{5}]=-k_{3},~[k_{2} , k_{3}]= -k_{2}, $ \\
  &      $k_{4}=  y \partial_{x}-x \partial_{y},$ &   $[k_{2} , k_{5}]=-k_{4}, [k_{2} , k_{4}]=k_{1}, $ \\
  &      $k_{5}= \partial_{x},$  & $[k_{2} , k_{6}]=-k_{3}, [k_{3} , k_{5}]=-k_{5},$ \\
  &      $k_{6}=   \partial_{y}$  & $[k_{3} , k_{6}]= -k_{6},  [k_{4} , k_{5}]= k_{6},  $ \\
    &        & $ [k_{4} , k_{6}]= -k_{5},  [k_{1} , k_{6}]= k_{4}$ \\\hline

 &       $k_{1}= -\cos z~  \partial_{y}+\sin z ~ \cot y~  \partial_{z}, $  &  $ {\bf so(3) \oplus u(1) }:$ \\
$E^1\times S^2$ &  $k_{2}= \sin z~  \partial_{y}+\cos z ~ \cot y~  \partial_{z},$ &  $~~~~[k_{1} , k_{2}]=k_{3},~[k_{3} , k_{1}]= k_{2},$\\
  &      $k_{3}= \partial_{z},~~~~~~~~k_{4}= \partial_{x}$ & $~~~~[k_{2} , k_{3}]=k_{1},~[k_{4} , .]=0$ \\\hline

   &       $k_{1}= e^{-z}\big(\partial_{y}+\tanh y~\partial_{z}\big), $  &                           ${\bf sl(2,\mathbb{R}) \oplus u(1) }: $ \\
$E^1\times H^2$ &  $k_{2}= e^{z}\big(-\partial_{y}+\tanh y~\partial_{z}\big),$ & $~~~[k_{1} , k_{2}]= 2k_{3},~ [k_{1} , k_{3}]= k_{1},$\\
  &      $k_{3}=  \partial_{z},~~~~~~~~~~~~k_{4}= \partial_{x}$ & $~~~[k_{3} , k_{2}]=k_{2},~[k_{4} , .]=0$ \\\hline

   & $k_{1}=  y \partial_{x} -x \partial_{y} -\frac{1}{2} (x^2-y^2)\partial_{z},$  &  ${\bf A_{4, 10} }:~ [k_{1} , k_{2}]=-k_{3},$ \\
Nil &    $k_{2}= \partial_{x} +y  \partial_{z},$ &  $~~~~~~~~~~~[k_{1} , k_{3}]=k_{2},$\\
  &      $k_{3}= - \partial_{y},$ & $~~~~~~~~~~~[k_{2} , k_{3}]=k_{4},$ \\
  &      $k_{4}=  \partial_{z}$ &  $~~~~~~~~~~~[k_{4} , .]=0$\\\hline

  &      $k_{1}= x y  \partial_{x}-\frac{1}{2} (x^2 -y^2) \partial_{y} +x\partial_{z}, $  & ${\bf sl(2,\mathbb{R}) \oplus u(1) }:$ \\
${\widetilde{SL}}(2 , \mathbb{R})$ &  $k_{2}= x \partial_{x}+y \partial_{y},$ &                   $~[k_{1} , k_{2}]=-k_{1},~[k_{2} , k_{3}]= -k_{3},$\\
  &                             $k_{3}= -\partial_{y},$ &                                   $~[k_{1} , k_{3}]=k_{2},~~~[k_{4} , .]=0$ \\
    &                           $k_{4}= \partial_{z}$ &                                       \\\hline

       &  $k_{1}= x \partial_{x} - y \partial_{y} +\partial_{z}, $  &  ${\bf VI_0 }:$ \\
Sol    &  $k_{2}= \partial_{y},$                                    & $~~~~~~[k_{1} , k_{2}]=k_{2},$\\
       &  $k_{3}= \partial_{x}$                                     & $~~~~~~[k_{1} , k_{3}]=- k_{3}$ \\\hline
		\end{tabular}}}
\end{center}


\begin{center}
		\scriptsize {{{\bf Table 4.}~ Killing vectors and isometry Lie algebras of Lorentzian Thurston geometries.}}
		{\scriptsize
			\renewcommand{\arraystretch}{1.5}{
\begin{tabular}{| l|l|l| } \hline \hline
Geometry  & Killing vectors & Lie algebra of  \\  &   &  isometry group\\ \hline
          &       $k_{1}= -( y \partial_{x}+x \partial_{y}), $  &  ${\bf {{\cal D} {\cal D}}_7: {(VII_0|IV|b)}_{_{b=1}}} $ \\
     &  $k_{2}=-(  z \partial_{x}+x \partial_{z}),$ &  $[k_{1} , k_{4}]=-k_{2},~[k_{1} , k_{5}]= -k_{3},$\\
$M^3$ &      $k_{3} =-\partial_{x},$ & $[k_{2} , k_{3}]=-k_{6},~ [k_{2} , k_{4}]=k_{1},$ \\
     &      $k_{4}=   z \partial_{y}-y \partial_{z},$ & $[k_{2} , k_{6}]=-k_{3}, ~[k_{4} , k_{5}]=k_{6},$ \\
     &    $k_{5} = \partial_{y},$  & $ [k_{4} , k_{6}]= -k_{5}, ~[k_{1} , k_{3}]= -k_{5}$ \\
          &    $k_{6}=   \partial_{z}$  & $[k_{1} , k_{2}]=-k_{4}$ \\\hline

   &       $k_{1}= -(\partial_{x}+y \partial_{z}), $  &                                               ${\bf A_{4, 10} }:$  \\
Nil &  $k_{2}= - \partial_{z},~~k_{3}=  \partial_{y},$ & $~~[k_{1} , k_{3}]=-k_{2},~ [k_{1} , k_{4}]=- k_{3},$\\
  &      $k_{4}=x \partial_{y} - y  \partial_{x} +\frac{1}{2}(x^2-y^2) \partial_{z},$ & $~~[k_{3} , k_{4}]=k_{1},~~[k_{2} , .]=0$ \\\hline

    &    $k_{1}= -\sin \Phi~ {\csc \Theta} ~\partial_{_\Psi}- {\cos \Phi}~  \partial_{_\Theta} +\sin \Phi~ {\cot \Theta} ~ \partial_{_\Phi}, $  &  ${\bf so(3) \oplus u(1)}: $ \\
   $SL(2 , \mathbb{R})$ &    $k_{2}= -\cos \Phi~ {\csc \Theta} ~\partial_{_\Psi}+ {\sin \Phi}~  \partial_{_\Theta} +\cos \Phi~ {\cot \Theta} ~ \partial_{_\Phi}, $  & $[k_{1} , k_{2}]=-k_{4}, ~[k_{1} , k_{4}]= k_2,$ \\
 & $k_{3}= -\partial_{_\Psi}, ~~~k_{4}=-\partial_{_\Phi} $  &   $[k_{2} , k_{4}]=-k_{1},~[k_{3} , .]=0$\\\hline

           &  $k_{1}= -x \partial_{x} + y \partial_{y} -\partial_{z}, $  &  ${\bf VI_0 }:$ \\
New Sol    &  $k_{2}= \partial_{y},$                                    & $[k_{1} , k_{2}]=-k_{2},~~[k_{1} , k_{3}]=k_{3}$\\
           &  $k_{3}= \partial_{x}$                                     &  \\\hline

 &       $k_{1}= x \partial_{x} -y  \partial_{y}+  \partial_{z}, $  &  $ {\bf {A^{b=-\frac{1}{2}}_{4,9}} }:$ \\
Lorentz Sol &  $k_{2}= \partial_{x},$ &  $~~~~[k_{1} , k_{2}]=-k_{2},~[k_{1} , k_{4}]= k_{4},$\\
  &      $k_{3}= y \partial_{x} +\frac{1}{3} e^{-3z} \partial_{y},~~~~k_{4}= \partial_{y}$ & $~~~~[k_{3} , k_{4}]=-k_{2},~[k_{1} , k_{3}]=-2k_{3}$ \\\hline

            &  $k_{1}= x \partial_{x} - y \partial_{y} +\partial_{z}, $  &  ${\bf VI_0 }:$ \\
Third Sol   &  $k_{2}= -\partial_{x},$                                    & $~~~[k_{1} , k_{2}]=-k_{2},$\\
            &  $k_{3}= -\partial_{y}$                                     & $~~~[k_{1} , k_{3}]= k_{3}$ \\\hline

   & $k_{1}=  \partial_{x} +y  \partial_{z},$  &  ${\bf A_{4, 8} }:$ \\
Lorentz-Heisenberg &    $k_{2}=  \partial_{z},$ &  $[k_{1} , k_{3}]=k_{2},$\\
  &      $k_{3}= - \partial_{y},$ &                $[k_{1} , k_{4}]=k_{3},$ \\
  &      $k_{4}=  -x \partial_{y} - y \partial_{x} -\frac{1}{2} (x^2 +y^2)\partial_{z},$ & $[k_{3} , k_{4}]=k_{1}$ \\\hline

		\end{tabular}}}
\end{center}
$\\\\$
Moreover, the isometry Lie algebra of both Euclidean and Lorentzian geometry of Nil is isomorphic to the $A_{4, 10}$ (Nappi-Witten Lie algebra) of Patera's classification \cite{patera}, and for the Lorentz Sol and Lorentz-Heisenberg we have found the isomorphism Lie algebras
${A^{b=-\frac{1}{2}}_{4,9}}$ and $A_{4, 8}$, respectively \cite{patera}.
We have also specified the label of isometry Lie algebras corresponding to the other geometries (see Tables 3 and 4).

In order to construct the dualizable metrics we need a non-Abelian three-dimensional Lie subalgebra
(Lie subalgebras of the Bianchi type) of the isometry Lie algebra that generates
the isometry subgroup acting freely and transitively on a three-dimensional target manifold where
the metrics of Thurston geometries are defined.
For the $E^1 \times S^2$ geometry we find that only the isometry Lie subalgebra is $IX \cong so(3)$, and
the action of corresponding isometry subgroup is not transitive on this geometry.
Also, for the $S^3$ geometry only the isometry Lie subalgebra is $IX$.

It can simply shown that the action of the corresponding isometry subgroup is not transitive in every point of the manifold $\M=S^3$, because
we find that $\det A_a^{~\mu} = 1/\sin x$. This means that the condition of transitivity \eqref{3.1} is not satisfied in every point of $\M$.
For geometry  $SL(2 , \mathbb{R})$ of Table 4 one can see that the result of the transitive action of isometry subgroup on the corresponding manifold
is the same. Hence, one cannot apply the non-Abelian T-duality approach described in section \ref{Sec.II} to obtain the target space duals of
the $E^1 \times S^2$, $S^3$ and $SL(2 , \mathbb{R})$ metrics.
The obtained results are summarized in Tables 5 and 6. All these results will be our main tools in the next section.\\
\\
\\
\begin{center}
		\scriptsize {{{\bf Table 5.} Isometry Lie subalgebras, and free and transitive action of the isometry subgroups on   \\
 Euclidean Thurston geometries.$~~~~~~~~~~~~~~~~~~~~~~~~~~~~~~~~~~~~~~~~~~~~~~~~~~~~~~~~~~~$}}
		{\scriptsize
			\renewcommand{\arraystretch}{1.5}{
\begin{tabular}{| l|l|l| l|} \hline \hline
Geometry  & Three-dimensional isometry Lie subalgebras & Free action  & Transitive action\\ \hline
        &       $VII_0:~Span\{ T_1 = -\alpha_0 k_6,~T_2 = \alpha_0 k_5, ~T_3 = k_3+ k_4\}$  & Yes & Yes \\
$E^3$   &       $IX:~Span\{ T_1 = k_{1}- \rho k_{3} +\lambda k_{5}, $  & Yes & No \\
        &       $~~~~~~~~~~~~~~~~T_2=k_{2}+\beta k_{3} +\lambda k_{6},$  &  & \\
       &       $~~~~~~~~~~~~~~~~T_3=k_{4}+\beta k_{5} +\rho k_{6}\}$  &  & \\\hline

$S^3$   &       $IX:~Span\{ T_1 = -k_3,~ T_2 = k_5,~T_3 =  k_4\}$  & Yes for   & Yes for  \\
        &        &  $\sin x \neq 0$  & $\sin x \neq 0$ \\\hline

        &        $V:~Span\{ T_1 = -k_3,~T_2 =  k_5,~T_3 = k_6\}$      & Yes & Yes \\
        &        $VII_0:~Span\{ T_1 = -k_2,~T_2 =  k_1,~T_3 = k_4\}$  & Yes & No \\
        &     $VII_a:~Span\{ T_1 = a k_3 +k_4,~T_2=k_{1}-  k_{2} $  & Yes & Yes \\
$H^3$   &           $~~~~~~~~~~~~~~~~~~~~~~~~~~~~~~~~~~~~~~T_3= k_{1}+ k_{2}\}$  &  & \\
        &           $VIII:~Span\{ T_1 = k_{2}+ k_{3},~T_2=k_{3}+k_{6}, $  & Yes & No \\
        &           $~~~~~~~~~~~~~~~~~~~~~~~~~~~~~~~~T_3=k_{2}+ k_{3} + k_{6}\}$  &  & \\
        &           $IX:~Span\{ T_1 = k_{1}+ k_{4},~T_2=-k_{4}+k_{5}, $  & Yes & No \\
        &           $~~~~~~~~~~~~~~~~~~~~~~~~~~~~~~~~T_3=k_{2}- k_{3} + k_{6}\}$  &  & \\\hline

  $E^1 \times S^2$  &  $IX:~Span\{ T_1 = k_1,~ T_2 = k_2,~T_3 =  k_3\}$  & Yes  & No \\\hline

                 &                  $III:~Span\{ T_1 =k_2 +k_3,~T_2 =  k_2,~T_3 =  k_4\}$  & Yes & Yes \\
 $E^1 \times H^2$  &   $VIII:~Span\{ T_1 = k_{1} +k_2 -2k_3, ~ T_2=k_{1},$  & Yes & No \\
              &                     $~~~~~~~~~~~~~~~~~~~~~~~~~~~~~~~~~~~~~~~~T_3=k_{1}- k_{3}\}$  &  & \\\hline

Nil         &                 $II:~Span\{ T_1 = k_4,~ T_2 = k_2,~T_3 =  k_3\}$  & Yes  & Yes \\ \hline

   ${\widetilde{SL}}(2 , \mathbb{R})$  &  $III:~Span\{ T_1 = - k_2,~T_2 = k_3, ~T_3=k_4\} $    & Yes & Yes \\
                                &   $VIII:~Span\{ T_1 = k_{2},~T_2 = k_1, ~T_3=k_3 \}$  & Yes & Yes \\\hline

Sol      &  $VI_0:~Span\{ T_1 = k_1,~T_2 = k_2,~T_3 =  k_3\}$  & Yes  & Yes \\\hline

		\end{tabular}}}
\end{center}

$\\
\\$

\begin{center}
		\scriptsize {{{ \bf Table 6.}~ Isometry Lie subalgebras, and free and transitive action of the isometry subgroups on Lorentzian \\
  Thurston geometries.$~~~~~~~~~~~~~~~~~~~~~~~~~~~~~~~~~~~~~~~~~~~~~~~~~~~~~~~~~~~~~~~~~~~~~~~~~~~~~~~~~~~~~~$}}
		{\scriptsize
			\renewcommand{\arraystretch}{1.5}{
\begin{tabular}{| l|l|l| l|} \hline \hline
Geometry  & Three-dimensional isometry Lie subalgebras & Free action  & Transitive action\\ \hline

            &     $II:~Span\{ T_1 = - k_{3}+k_{5}, ~ T_2 =  k_6,~T_3 = -k_2+k_{4} +k_{5}\} $  & Yes & Yes \\
           &   $III:~Span\{ T_1 =  -k_{2},  ~T_2 =  k_{3} + k_{6}, ~T_3 =  \lambda_{_0} k_{5}\} $  & Yes & Yes \\
           &      $IV:~Span\{ T_1 =  k_{2} +k_{5},~T_2=k_{1}+k_{4}, ~ T_3=k_{3}-k_{6}\} $  & Yes & Yes \\
  $M^3$   &      $V:~Span\{ T_1 = - k_{1},~T_2=k_{2}+k_{4}, ~ T_3=k_{3}+k_{5}\} $  & Yes & Yes \\
           &      $VI_0:~Span\{ T_1 = k_{3},~T_2=k_{5}, ~ T_3=k_{1}+\gamma k_{6}\} $  & Yes & Yes \\
          &      $VII_0:~Span\{ T_1 = \sqrt{\lambda^2-1} ~ k_{3} +\lambda k_{5},~T_2=k_{6},$  &  &  \\
      &      $~~~~~~~~~~~~~~~~~~~~~~~~~~~~T_3=\sqrt{\lambda^2-1} ~ k_{2} +\lambda k_{4}+ \sigma k_{5}\} $  & Yes & Yes \\
          &      $VIII:~Span\{ T_1 = - k_{1} +\rho k_{3}-\sqrt{\rho^2-1} ~k_{5},$ & & \\
         & $~~~T_2=-\rho k_{2}+\sqrt{\rho^2-1} ~k_{4}, ~ T_3=\rho k_{4}+k_{6}- \sqrt{\rho^2-1}~ k_{2}\}$  & Yes & No \\\hline

Nil  &       $II:~Span\{ T_1 = -k_2,~T_2 = k_1, ~T_3 = k_3\}$  & Yes & Yes  \\\hline

$SL(2, \mathbb{R})$   &       $IX:~Span\{ T_1 = k_2,~ T_2 = k_1,~T_3 =  k_4\}$  & Yes for   & Yes for \\
                                            &       &  $\sin {\Theta}\neq0 $  &  $\sin {\Theta}\neq0 $ \\\hline

New Sol   &        $VI_0:~Span\{ T_1 = k_1,~T_2 =  k_2,~T_3 = k_3\}$      & Yes & Yes \\\hline

            &     $II:~Span\{ T_1 =  k_{2}, ~ T_2 =  k_4,~T_3 = k_3\} $  & Yes & No \\
Lorentz Sol &     $VI_0:~Span\{ T_1 =  k_{2} - k_{4},  ~T_2 =  k_{2} + k_{4}, ~T_3 =  k_{1}\} $  & Yes & Yes \\
           &      $VI_{_{a=3}}:~Span\{ T_1 = -2 k_{1},~T_2=k_{2}-k_{3}, ~ T_3=k_{2}+k_{3}\} $  & Yes & Yes \\\hline

Third Sol  &          $VI_0:~Span\{ T_1 = k_{1}, ~ T_2=k_{2}, ~ T_3=k_{3}\}$  & Yes & Yes \\\hline

Lorentz-Heisenberg  &  $II:~Span\{ T_1 = -k_2,~ T_2 = k_3,~T_3 =  k_1 +\alpha k_{2} + \beta k_{3}\}$  & Yes  & Yes \\ \hline

		\end{tabular}}}
\end{center}


\section{\label{Sec.IV} Non-Abelian target space duals of Thurston geometries}

Let us turn into the main goal of this paper which is nothing but calculating the
non-Abelian target space duals of Thurston geometries.
In what follows we apply the dualization procedure described in section \ref{Sec.II}
to investigate the metrics of Thurston geometries from the point of view
of their non-Ablelian T-dualizability (here as PL T-duality on a semi-Abelian double).
For that purpose we reconstruct the metrics as backgrounds of non-linear $\sigma$-models on Lie
groups of the Bianchi type. For construction of dual backgrounds we use Drinfeld doubles
obtained from the non-Abelian isometry subgroups of the metrics.
Note that the Drinfeld doubles for a $\sigma$-model living in curved background can sometimes be found
from the knowledge of symmetry group of the metric. As mentioned in the Introduction,
in case that the metric has sufficient number of independent Killing vectors (the same as the metrics of Thurston geometries),
the non-Abelian isometry subgroup of the metric can be taken as one of the subgroups of the Drinfeld double.
In order to satisfy the conditions of dualizability, the other one then must be chosen to be Abelian.

In our case, the Lie algebra $\D$ of the Drinfeld double can
be composed from the non-Abelian Bianchi Lie algebra $\G$  isomorphic to the
three-dimensional subalgebra of Killing vectors and three-dimensional Abelian
algebra ${\tilde \G}=I$. Most importantly, the three-dimensional subgroup of isometries must act
freely and transitively on the target manifold $\M$ where the metrics of Thurston geometries are defined so that $\M \approx G$.
As an example, below we discuss in details the non-Abelian T-dualization of the Lorentz Sol geometry.

\subsection{ An example: non-Abelian T-dualization of Lorentz Sol geometry}

In this subsection, we shall investigate the non-Abelian T-duality of the metric of Lorentz Sol geometry by the PL T-duality approach.
For that purpose, one must obtain the isometry subgroups of this metric acting
freely and transitively on the corresponding manifold (Table 6).
According to Table 2, Lorentz Sol geometry is defined by metric
\begin{eqnarray}
ds^2=l^2 \Big[2 e^{-z} dx~dz+e^{2 z} dy^2\Big].\label{4.1}
\end{eqnarray}
This metric has a number of symmetries important for construction of
the dualizable $\sigma$-models. It admits the following Killing vectors
\begin{eqnarray}
	k_{_1}&=& x \partial_{_x}-y \partial_{_y} +\partial_{_z},\nonumber\\
	k_{_2}&=&\partial_{_x},\nonumber\\
	k_{_3}&=&y \partial_{_x}+\frac{1}{3} e^{-3z} \partial_{_y},\nonumber\\
	k_{_4}&=&\partial_{_y}.\label{4.2}
\end{eqnarray}
The isometry Lie algebra spanned by these vectors is
\begin{eqnarray}\label{4.3}
[k_{1} , k_{2}]=-k_{2},~~~~[k_{1} , k_{3}]=-2 k_{3},~~~~[k_{1} , k_{4}]=k_{4},~~~~~~[k_{3} ,k_{4}]=-k_{2}.
\end{eqnarray}
One can easily check that the Lie algebra spanned by these Killing vectors is isomorphic to the $A_{4,9}^{b=-1/2}$ Lie algebra
of Patera's classification \cite{patera}.
There are three classes of three-dimensional subalgebras
of the isometry Lie algebra isomorphic to one of the following algebras:
\begin{eqnarray}
II:&&Span\big\{T_1=k_{2},~T_2=k_{4},~T_3=k_{3}\big\},\label{4.4}\\
VI_{_0}:&&Span\big\{T_1=k_{2}-k_{4},~T_2=k_{2}+k_{4},~T_3=k_{1}\big\},\label{4.5}\\
VI_{_{a=3}}:&&Span\big\{T_1=-2 k_{1},~T_2=k_{2}-k_{3},~T_3=k_{2} +k_{3}\big\}. \label{4.6}
\end{eqnarray}
By inspection we can find that the only three-dimensional subalgebras  that
generate transitive action on the manifold defined by Lorentz Sol are isomorphic to $VI_{_0}$ and $VI_{_{a=3}}$.
One can also check that the action of both corresponding groups of isometries is free.
In the following we shall find metrics dual to \eqref{4.1} that follow from its Drinfeld double description
where $\G$ is isomorphic either to algebra spanned by $VI_{_0}$ or by $VI_{_{a=3}}$.

\subsubsection{ Dualization with respect to subgroup generated by $VI_{_0}$}
Let us start with construction of the Drinfeld double following from the
algebra isomorphic to \eqref{4.5} and dual Abelian algebra. The Lie
algebra $\G=VI_{_0}$ is spanned by elements $(T_1 , T_2, T_3)$ with commutation relations
\begin{eqnarray}\label{4.7}
[T_2 , T_3]=T_1,~~~~~~[T_3 , T_1]=-T_2.
\end{eqnarray}
In order to obtain the left-invariant vector fields of the group generated by $VI_{_0}$, we use the following parametrization
of the group manifold:
\begin{eqnarray}\label{4.8}
g~=~ e^{x_3 T_3} ~e^{x_1 T_1}~e^{x_2 T_2},
\end{eqnarray}
where $(x_1, x_2, x_3)$ are group coordinates. By using \eqref{4.7} and \eqref{4.8} one can find the left-invariant one-form
on $VI_{_0}$
\begin{eqnarray}\label{4.9}
g^{-1} dg~=~(dx_1 -x_2 dx_3) T_1 + (dx_2-x_1 dx_3) T_2 +dx_3  T_3,
\end{eqnarray}
from which we can read off the basis of left-invariant vector fields
\begin{eqnarray}\label{4.10}
V_1 &=& \partial_{_{x_1}},\nonumber\\
V_2 &=& \partial_{_{x_2}},\nonumber\\
V_3 &=& x_2 \partial_{_{x_1}} +x_1 \partial_{_{x_2}} +\partial_{_{x_3}}.
\end{eqnarray}
To be able to obtain the metric \eqref{4.1} by the Drinfeld double construction first
we have to transform it into the group coordinates. By converting the Killing vectors $T_1=k_{2}-k_{4},~T_2=k_{2}+k_{4},~T_3=k_{1}$
into the left-invariant vector fields, one can get transformation between
group coordinates $(x_1, x_2, x_3)$ and geometrical coordinates $(x, y, z)$, giving us
\begin{eqnarray}\label{4.11}
x &=& x_1 +x_2 +e^{x_3},\nonumber\\
y &=&-x_1 +x_2 +e^{-x_3},\nonumber\\
z &=& x_3.
\end{eqnarray}
This transformation converts \eqref{4.1} into the new form of line element on the group generated by $VI_{_0}$
\begin{eqnarray}\label{4.12}
ds^2&=&l^2\Big[e^{2x_3} dx_1^2 +e^{2x_3} dx_2^2 +3 dx_3^2 -2 e^{2x_3} dx_1~dx_2\nonumber\\
&&~~~~~~~~~~~~~~~~~~~~~+ 4 \cosh x_3~ dx_1~dx_3-4\sinh x_3~ dx_2~dx_3\Big],
\end{eqnarray}
from which one can write the metric $G_{\mu \nu}$. Since the $B$-field is absent, the corresponding background matrix is then expressed as
\begin{eqnarray}\label{4.13}
{\cal E}_{\mu \nu}= l^2 \left(
\begin{array}{ccc}
	e^{2x_3} & -e^{2x_3} &  2\cosh x_3 \\
	-e^{2x_3} & e^{2x_3} &  -2 \sinh x_3 \\
    2 \cosh x_3 & -2 \sinh x_3 &    3\\
\end{array}
\right).
\end{eqnarray}
To get the matrix $E_0$ necessary for construction of the dual model we note that it is given by the value of $E(g)$
in $g=e$, i.e. by value of ${\cal E}_{\mu \nu}$ for $x_i=0, i=1, 2, 3$.
Comparing \eqref{4.13} with \eqref{sec2.14} and then using the fact that ${R_{\mu}^{~a}}_{|_{x^\mu =0}} = \delta_{\mu}^{~a}$, we get
\begin{eqnarray}\label{4.14}
E_0(e)= l^2 \left(
\begin{array}{ccc}
	1 & -1 &  2 \\
	-1 & 1 & 0 \\
    2  & 0 &  3\\
\end{array}
\right).
\end{eqnarray}
The dual background on the Abelian group $\tilde G$ can be constructed by the procedure explained in section \ref{Sec.II},
namely by using equations \eqref{sec2.11} and \eqref{sec2.12}. First of all, we need to form the Drinfeld double generated by
the Lie algebra $VI_{_0}$ and its dual pair, the Bianchi type algebra $I$ spanned by the set of generators
$({\tilde T}^{1}, {\tilde T}^{2}, {\tilde T}^{3})$. The Lie algebra of the semi-Abelian double
$(VI_{_0}~,~ I)$ obeys the following set of non-trivial commutation relations \cite{{JR},{Hlavaty2}}
\begin{eqnarray}
[T_{2} , T_{3}] &=&T_{1},~~~~~~~~~[T_{1} , T_{3}]=T_{2},~~~~~~~~~[T_{1} , {\tilde T}^{2}]=-{\tilde T}^{3},\nonumber\\
{[T_{2} , {\tilde T}^{1}]}&=&-{\tilde T}^{3},~~~~~~[T_{3} , {\tilde T}^{1}]={\tilde T}^{2},~~~~~~~~
[T_{3} , {\tilde T}^{2}]={\tilde T}^{1}.\label{4.15}
\end{eqnarray}
As mentioned in section \ref{Sec.II}, to obtain the dual Poisson structure one must use equation \eqref{sec2.10} and
then second formula of \eqref{sec2.9}
by replacing untilded quantities with tilded ones. Thus, we obtain
\begin{eqnarray}\label{4.16}
{\tilde \Pi}_{ab}(\tilde g)=\left(
\begin{array}{ccc}
	0 & 0 &  -{\tilde x}_2 \\
	0 & 0 & -{\tilde x}_1 \\
   {\tilde x}_2  & {\tilde x}_1 &  0\\
\end{array}
\right).
\end{eqnarray}
Here, the dual group coordinates have been considered to be $({\tilde x}_1, {\tilde x}_2, {\tilde x}_3)$.
Noting the fact that the dual Lie algebra is Abelian we find that $\tilde R_{_{\mu a}}=\delta_{_{\mu a}}$.
Finally, inserting \eqref{4.16} and \eqref{4.14} into \eqref{sec2.12} and then using \eqref{sec2.11},
the dual action is worked out
\begin{eqnarray}
{\tilde S}&=&\frac{1}{2}\int d\sigma^+d\sigma^- \Big\{\frac{1}{l^2 \Delta_{_1}}\big[-(3l^4+\tilde{x}_1^2) \partial_+\tilde x_1 \partial_-\tilde x_1
+(l^4-\tilde{x}_2^2) \partial_+\tilde x_2\partial_-\tilde x_2~~ \nonumber\\
&&~~~~~~~~~~-(3l^4+2l^2 \tilde{x}_1-\tilde{x}_1 \tilde{x}_2) \partial_+\tilde x_1 \partial_-\tilde x_2-(3l^4-2l^2 \tilde{x}_1-\tilde{x}_1 \tilde{x}_2) \partial_+\tilde x_2 \partial_-\tilde x_1\big]~~\nonumber\\
&&~~~~~~~~~~~~~~~~~~~~+\frac{(\partial_+\tilde x_1 \partial_-\tilde x_3+\partial_+\tilde x_2 \partial_-\tilde x_3)}{2 l^2 +\tilde{x}_1+\tilde{x}_2}
+\frac{(\partial_+\tilde x_3 \partial_-\tilde x_1+\partial_+\tilde x_3 \partial_-\tilde x_2)}{2 l^2 -\tilde{x}_1-\tilde{x}_2}\Big\},\label{4.17}
\end{eqnarray}
where $\Delta_{_1}=4l^4-(\tilde{x}_1+\tilde{x}_2)^2$.
One may compare the dual model \eqref{4.17} with dual version of general $ \sigma $-model \eqref{sec2.2} to obtain
\begin{eqnarray}
d\tilde s^2&=&\frac{1}{l^2 \Delta_{_1}}\Big[-(3l^4+\tilde{x}_1^2)d\tilde x_1^2+(l^4-\tilde{x}_2^2) d\tilde x_2^2\nonumber\\
&&~~~~~~~~~~~~~~~~~~-2(3l^4-\tilde{x}_1 \tilde{x}_2) d\tilde x_1 d\tilde x_2 +4 l^4 (d\tilde x_1 d\tilde x_3 + d\tilde x_2 d\tilde x_3)\Big],\label{4.18}\\
\tilde B&=&\frac{1}{\Delta_{_1}} \Big[-2 \tilde x_1~ d\tilde x_1\wedge d\tilde x_2 - (\tilde{x}_1+\tilde{x}_2)
(d\tilde x_1\wedge d\tilde x_3+ d\tilde x_2\wedge d\tilde x_3)\Big].\label{4.19}
\end{eqnarray}

\subsubsection{ Dualization with respect to subgroup generated by $VI_{_{a=3}}$}
Here, for dualization of the metric \eqref{4.1} we shall use the Lie algebra $\G=VI_{_{a=3}}$ spanned by
$(T_1=-2 k_{1},~T_2=k_{2}-k_{3},~T_3=k_{2} +k_{3})$ with non-zero commutation relations
\begin{eqnarray}\label{4.20}
[T_1 , T_2]=3T_2 - T_3,~~~~~[T_3 , T_1]=T_2 -3T_3.
\end{eqnarray}
Note that the corresponding Drinfeld double is generated by the Lie algebra $VI_{_{a=3}}$ defined
by the commutation relations \eqref{4.20} and three-dimensional Abelian algebra or the same the Bianchi type algebra $I$.
The basis of left-invariant vector fields of the group generated by $VI_{_{a=3}}$ is
\begin{eqnarray}\label{4.21}
V_1 = \partial_{_{x_1}} + (x_3-3 x_2) \partial_{_{x_2}}+ (x_2-3 x_3) \partial_{_{x_3}},~~~~~~V_2 = \partial_{_{x_2}},~~~~~~~
V_3 = \partial_{_{x_3}},
\end{eqnarray}
where $(x_1, x_2, x_3)$ are group coordinates used in parametrization
\begin{eqnarray}\label{4.22}
g~= ~e^{x_1 T_1}~e^{x_2 T_2}~e^{x_3 T_3}.
\end{eqnarray}
Transformation between the group coordinates and coordinates $(x, y, z)$ of the Lorentz Sol manifold is
\begin{eqnarray}\label{4.23}
x &=& \frac{1}{6} e^{6 x_1} (x_2 -x_3)^2 +x_2 +x_3,\nonumber\\
y &=&-\frac{1}{3} e^{6 x_1} (x_2-x_3),\nonumber\\
z &=& -2 x_1.
\end{eqnarray}
This transformation converts the Killing vectors $(k_{1}, k_{2}, k_{3})$ into
the left-invariant vector fields \eqref{4.21} and thus the metric \eqref{4.1} into
\begin{eqnarray}\label{4.24}
{\cal E}_{\mu \nu}={G}_{\mu \nu}= l^2 \left(
\begin{array}{ccc}
	0 & -2 e^{2x_1} &  -2 e^{2x_1} \\
	-2 e^{2x_1} & \frac{1}{9} e^{8 x_1} &  -\frac{1}{9} e^{8x_1} \\
    -2 e^{2x_1} & -\frac{1}{9} e^{8x_1} &    \frac{1}{9} e^{8x_1}\\
\end{array}
\right).
\end{eqnarray}
The value of this metric for $x_i=0, i=1, 2, 3$, i.e. in the unit of the
group, gives the $\sigma$-model constant matrix in the following form
\begin{eqnarray}\label{4.25}
E_0(e)= l^2 \left(
\begin{array}{ccc}
	0 & -2 &  -2 \\
	-2 & \frac{1}{9} & -\frac{1}{9} \\
    -2  & -\frac{1}{9} & \frac{1}{9}\\
\end{array}
\right).
\end{eqnarray}
Having this matrix one can construct the action of dual model.
It is again obtained using \eqref{sec2.11} and \eqref{sec2.12} and has the form
\begin{eqnarray}
{\tilde S}&=&\frac{1}{2}\int d\sigma^+d\sigma^- \Big\{\frac{9}{4 l^2 \Delta_{_2}}\big[\delta_{_+} \delta_{_-} \partial_+\tilde x_2 \partial_-\tilde x_2+ \lambda_{_+} \lambda_{_-} \partial_+\tilde x_3 \partial_-\tilde x_3 -\delta_{_-} \lambda_{_-} \partial_+\tilde x_2 \partial_-\tilde x_3~~~~\nonumber\\
&&  -\delta_{_+} \lambda_{_+} \partial_+\tilde x_3 \partial_-\tilde x_2\big]-\frac{(\partial_+\tilde x_1 \partial_-\tilde x_2+\partial_+\tilde x_1 \partial_-\tilde x_3)}{2(2 l^2 -\tilde{x}_2-\tilde{x}_3)} -\frac{(\partial_+\tilde x_2 \partial_-\tilde x_1+\partial_+\tilde x_3 \partial_-\tilde x_1)}
{2(2 l^2 +\tilde{x}_2+\tilde{x}_3)}\Big\},~~~~~\label{4.26}
\end{eqnarray}
where $\Delta_{_2} = 4l^4-(\tilde{x}_2 + \tilde{x}_3)^2, \delta_{_{\pm}}=2l^2 \pm (\tilde{x}_2 -3 \tilde{x}_3)$ and
$\lambda_{_{\pm}}=2l^2 \pm (3\tilde{x}_2 - \tilde{x}_3)$.
Finally, background including the symmetric metric,
the anti-symmetric tensor field is, in the coordinate basis, read off
\begin{eqnarray}
d\tilde s^2&=&\frac{9}{4 l^2 \Delta_{_2}}\Big[\delta_{_{+}} \delta_{_{-}} d\tilde x_2^2 +\lambda_{_{+}} \lambda_{_{-}} d\tilde x_3^2
-2\big(4l^4+3 (\tilde{x}_2^2 + \tilde{x}_3^2) -10\tilde{x}_2\tilde{x}_3\big)  d\tilde x_2  d\tilde x_3\Big]\nonumber\\
&&~~~~~~~~~~~~~~~~~~~~~~~~~~~~~~~~~~~~~~~~~~~~~~~~~~~~~-\frac{2 l^2}{\Delta_{_2}}\big(d\tilde x_1 d\tilde x_2 +d\tilde x_1 d\tilde x_3\big),\label{4.27}\\
\tilde B&=&-\frac{(\tilde{x}_2 + \tilde{x}_3)}{2 \Delta_{_2}}\big(d\tilde x_1\wedge d\tilde x_2 +d\tilde x_1\wedge d\tilde x_3\big)
+\frac{18(\tilde{x}_2 - \tilde{x}_3)}{\Delta_{_2}}~d\tilde x_2\wedge d\tilde x_3.\label{4.28}
\end{eqnarray}

In summary, the Lorentz Sol metric possessed isometry group whose dimension was greater than
the dimension of the manifold; moreover, it admited three isometry subgroups in a way that the action of two of them on the manifold was free and transitive.
Therefore, we could construct two backgrounds dual to the metric.
Similarly to this example, we obtain non-Abelian T-dual spaces for the remaining Euclidean and Lorentzian metrics.
For the sake of clarity the results obtained in this section are summarized in Tables 7 and 8;
we display the constant matrix $E_0(e)$, the transformation between the coordinates of Thurston metrics and group ones, together with the
metrics and $B$-fields corresponding to both original and dual backgrounds.

Before closing this subsection, let us comment on
the non-Abelian T-dualization of the Nil geometry previously discussed in Ref. \cite{bugden}.
There, it has been shown that there is a class of non-Abelian T-dualities that can be interpreted instead as a chain of Abelian T-dualities. In addition, the relationship between arbitrary non-Abelian T-duals and T-folds has been discussed.
First of all, a three-torus $\mathbb{T}^3$ with metric $ds^2 =dx^2+dy^2+dz^2$ (metric $E^3$ from Table 1)
and $B$-field $B=-x dy \wedge dz$ are considered.
Then, by performing  an Abelian T-duality along the vector $\partial_{_z}$, a dual space known as the twisted
torus is obtained, which is given by the following metric and $B$-field,
\begin{align*}
ds^2 &=dx^2+dy^2+(d\hat{z}-xdy)^2,\\
B&=0.
\end{align*}
This metric displays the Nil geometry which is a non-trivial circle bundle over $\mathbb{T}^2$.
Finally, by using the fact that $\partial_{_y}$ is a Killing vector of the Nil geometry,
one can therefore perform another T-duality along it. The result is the
so-called Q-flux background which is equivalent to the non-Abelian dual solution for the Nil geometry that we have presented in Table 7.
\\
\\
\\
\begin{center}
	\small {{{\bf Table 7.}~ Non-Abelian T-dual spaces of Euclidean Thurston geometries.}}
	{\scriptsize
		\renewcommand{\arraystretch}{1.5}{
			\begin{tabular}{| l|l|l| l| l|} \hline \hline
				Geometry${ \large \diagup}$  & Constant matrix & Coordinate & Original& Dual \\
				{\scriptsize Isometry subgroup} & $E_{_0}(e)$ & transformation &$\sigma$-model & $\sigma$-model \\ \hline

				\multirow{5}{*}{$E^3{ \large \diagup} VII_0$} & \multirow{5}{*}{$\left(\begin{array}{ccc}
				 		\alpha_0^2 & 0 &  0 \\
				 		0 & \alpha_0^2 &  0 \\
				 		0 & 0 &    1\\
				 	\end{array} \right)$}    & $x=x_3$ & $ ds^2= \alpha_0^2 \big[dx_1^2 +dx_2^2$  &  $d{\tilde s}^2= \frac{1}{\alpha_0^2 \Delta} \big[({\tilde x}_1^2 + \alpha_0^2 ) d{\tilde x}_1^2 $ \\

				  &  & $y=\alpha_0 [-x_1 \sin x_3 $ & $ +2 dx_3 (x_2 dx_1 -  x_1dx_2)\big]$  &  $+ ({\tilde x}_2^2 + \alpha_0^2 ) d{\tilde x}_2^2 +\alpha_0^4 d{\tilde x}_3^2 $ \\
				
				&    &  $+x_2 \cos x_3]$ & $ +(1+\alpha_0^2 x_1^2 +\alpha_0^2 x_2^2) dx_3^2, $  &  $+ 2 {\tilde x}_1 {\tilde x}_2  d{\tilde x}_1 d{\tilde x}_2\big], $ \\
				
				&    & $z=-\alpha_0 [x_1 \cos x_3$ &  $ B=0$  &  ${\tilde B}= \frac{1}{\Delta} \big[{\tilde x}_2  d{\tilde x}_3\wedge  d{\tilde x}_1 $ \\
				
				&    &     $+x_2 \sin x_3]$   &   &  $+{\tilde x}_1  d{\tilde x}_2 \wedge  d{\tilde x}_3\big]$ \\\hline
				
				\multirow{4}{*}{$E^1 \times H^2{ \large \diagup} III$  }&\multirow{4}{*}{$\left(\begin{array}{ccc}
						l^2 & 0 &  0 \\
						0 & l^2 &  0 \\
						0 & 0 &    l^2\\
					\end{array} \right)$} &
$x_1=z +\log(\cosh y),$  &  $ds^2= l^2 \big[dx_1^2 +dx_3^2$&  $d{\tilde s}^2= \frac{l^2}{l^4 +{\tilde x}_2^2} \big(d{\tilde x}_1^2  + d{\tilde x}_2^2 \big)$ \\
&  &  $x_2=-(\sinh y$    &  $+(x_2  dx_1- dx_2)^2\big], $ &  $~~~~+ ~\frac{d{\tilde x}_3^2}{l^2},$ \\
&  &  $+e^x \cosh y),$   &   &   \\
&  &  $x_3 = x$          &  $B=0$  &  ${\tilde B}=  \frac{{\tilde x}_2}{{l^4 +{\tilde x}_2^2}} d{\tilde x}_1 \wedge  d{\tilde x}_2 $ \\\hline

	\end{tabular}}}
\end{center}

\begin{center}
	\small {{{\bf Table 7.}~ Continued.}}
	{\scriptsize
		\renewcommand{\arraystretch}{1.5}{
			\begin{tabular}{| l|l|l| l| l|} \hline \hline
				Geometry${ \large \diagup}$  & Constant matrix & Coordinate & Original& Dual \\
				{\scriptsize Isometry subgroup} & $E_{_0}(e)$ & transformation &$\sigma$-model & $\sigma$-model \\ \hline
				
				\multirow{5}{*}{$H^3 {\Big / } V$}&\multirow{5}{*}{ $\left(\begin{array}{ccc}
						l^2 & 0 &  0 \\
						0 & l^2 &  0 \\
						0 & 0 &    l^2\\
					\end{array} \right)$}    &   $x=x_2 e^{-x_1}, $  &  $ds^2= l^2\big[dx_2^2 + dx_3^2$ &  $d{\tilde s}^2= \frac{1}{l^2 \Gamma} \big[(l^4 + {\tilde x}_3^2) d{\tilde x}_2^2 $ \\
				  & & $y=x_3 e^{-x_1},$  &  $-2  dx_1 (x_2 dx_2+ x_3 dx_3)$ &  $+ l^4 d{\tilde x}_1^2+(l^4 + {\tilde x}_2^2) d{\tilde x}_3^2 $ \\
				&  & $z= e^{-x_1}$ & $+(1+ x_2^2 + x_3^2) dx_1^2\big],$  &  $-2 {\tilde x}_2 {\tilde x}_3  d{\tilde x}_2 d{\tilde x}_3\big], $ \\
				&  &               &  $B=0.$              & ${\tilde B}=  \frac{1}{\Gamma} \big[{\tilde x}_2  d{\tilde x}_1 \wedge  d{\tilde x}_2 $  \\
				&  &              &                     &  $+{\tilde x}_3  d{\tilde x}_1 \wedge  d{\tilde x}_3\big]. $ \\
				&  &              &                     &   \\
				
				\multirow{5}{*}{$H^3 {\Big / } VII_{_a}$}&\multirow{5}{*}{ $l^2\left(\begin{array}{ccc}
						a^2 & 0 &  0 \\
						0 & \frac{1}{2} &  0 \\
						0 & 0 &   \frac{1}{2}\\
					\end{array} \right)$}    &   $x=-\frac{x_2 +x_3}{W}, $  &  $ds^2= l^2 [a^2 dx_1^2$ &  $d{\tilde s}^2= \frac{1}{\Xi} \big[(l^2 d{\tilde x}_1^2 $ \\
				  & & $y=\frac{x_2 -x_3}{W},$  &  $+\frac{1}{2}  e^{2 a x_1} (dx_2^2+ dx_3^2)\big],$ &  $+  \frac{2}{l^2}(\delta_1   d{\tilde x}_2^2+ \delta_2 d{\tilde x}_3^2) $ \\
				&  & $z= \frac{2 e^{-a x_1}}{W}$ &                 &  $-\frac{8}{l^2} \delta_3 d{\tilde x}_2 d{\tilde x}_3\big], $ \\
				&  &               &         $B=0$       & ${\tilde B}=  \frac{2}{\Xi} \big[(a{\tilde x}_2+{\tilde x}_3)  d{\tilde x}_1 \wedge  d{\tilde x}_2 $  \\
				&  &              &                     &  $+(a{\tilde x}_3 -{\tilde x}_2) d{\tilde x}_1 \wedge  d{\tilde x}_3\big] $ \\\hline

				\multirow{4}{*}{Nil ${ \large \diagup} II$  }&\multirow{4}{*}{$\left(\begin{array}{ccc}
						\frac{l^2}{l^4} & 0 &  0 \\
						0 & \frac{l^2}{l^4} &  0 \\
						0 & 0 &    \frac{l^2}{l^4}\\
					\end{array} \right)$} &
$x= x_2,$  &  $ds^2= \frac{l^2}{l^4} \big[dx_1^2 +dx_3^2$&  $d{\tilde s}^2= \frac{4}{l^2} d{\tilde x}_1^2$ \\
&  &  $y=-x_3,$    &  $+(1+x_3^2)dx_2^2  $ &  $+ ~\frac{l^2}{(\frac{l^4}{4} +4 {\tilde x}_1^2)}(d{\tilde x}_2^2 +d{\tilde x}_3^2),$ \\
&  &  $z=x_1-x_2 x_3$  & $-2x_3 d x_1 dx_2\big],$  &  \\
&  &            &  $B=0$   &  ${\tilde B}=  \frac{{\tilde x}_1}{(\frac{l^4}{16} + {\tilde x}_1^2)} d{\tilde x}_2 \wedge  d{\tilde x}_3 $  \\\hline

				\multirow{4}{*}{${\widetilde{SL}}(2 , \mathbb{R}) { \large \diagup} III$  }&\multirow{4}{*}{$\left(\begin{array}{ccc}
						1 & 0 &  0 \\
						0 & 2 &  -1 \\
						0 & -1 &   1\\
					\end{array} \right)$} &
$x= e^{-x_1},$  &  $ds^2=dx_3^2 -2d x_2 dx_3$&  $d{\tilde s}^2= \frac{1}{1+{\tilde x}_2^2}\big[d{\tilde x}_1^2 +2 d{\tilde x}_2 d{\tilde x}_3$ \\
&  &  $y=-x_2 e^{-x_1},$    &  $+2dx_2^2 +(1+2 x_2^2) dx_1^2 $ & $~~~+ (2+{\tilde x}_2^2) d{\tilde x}_3^2 +  d{\tilde x}_2^2\big],$ \\
&  &  $z=x_3$    & $-2 x_2 d x_1(2dx_2 -dx_3),$  &  ${\tilde B}=  \frac{{\tilde x}_2}{1+{\tilde x}_2} \big(d{\tilde x}_1 \wedge  d{\tilde x}_2 $ \\
&  &            &  $B=0.$  & $~~~~~~+ d{\tilde x}_1 \wedge  d{\tilde x}_3\big). $  \\
&  &            &   &   \\

				\multirow{4}{*}{${\widetilde{SL}}(2 , \mathbb{R}) { \large \diagup} VIII$  }&\multirow{4}{*}{$\left(\begin{array}{ccc}
						1 & 0 &  0 \\
						0 & \frac{1}{2} &  0 \\
						0 & 0 &   2\\
					\end{array} \right)$} &
$x= \frac{4 e^{x_1}}{4+x_2^2},$  &  $ds^2=\frac{(x_2^2+2)}{2} dx_1^2 +\frac{1}{2} dx_2^2$&  $d{\tilde s}^2= \frac{1}{\Sigma}\big[2(1+{\tilde x}_1^2)d{\tilde x}_1^2$ \\
&  &  $y=-x_3- \frac{2 x_2 e^{x_1}}{4+x_2^2},$    &  $+\frac{(x_2^2+4)^2}{8} e^{-2x_1} dx_3^2 $ & $+2(2+{\tilde x}_3^2)d{\tilde x}_2^2$ \\
&  &  $z=2 \arctan(\frac{x_2}{2})$    & $- x_2 d x_2(dx_1 +\frac{x_2 e^{-x_1}}{2} dx_3)$  &  $+(1+2{\tilde x}_2^2)d{\tilde x}_3^2$ \\
&  &            &  $+\frac{x_2 (x_2^2+4) e^{-x_1}}{2}  dx_1 dx_3,$  & $+4{\tilde x}_1 d{\tilde x}_1({\tilde x}_3 d{\tilde x}_2 +{\tilde x}_2 d{\tilde x}_3)$  \\
&  &            &    & $+4{\tilde x}_2  {\tilde x}_3 d{\tilde x}_2  d{\tilde x}_3\big],$  \\
&  &            &  $B=0$  & ${\tilde B}= \frac{1}{\Sigma}\big[4 {\tilde x}_2 d{\tilde x}_1 \wedge  d{\tilde x}_2$  \\
&  &            &   & $ -({\tilde x}_3 d{\tilde x}_1 - 2 {\tilde x}_1 d {\tilde x}_2)\wedge  d{\tilde x}_3\big]$  \\\hline

				\multirow{4}{*}{Sol $ { \large \diagup} VI_0$  }&\multirow{4}{*}{$\left(\begin{array}{ccc}
						l^2 & 0 &  0 \\
						0 & l^2 &  0 \\
						0 & 0 &    l^2\\
					\end{array} \right)$}
&   $x=x_3 e^{x_1}, $  &  $ds^2= l^2\big[dx_2^2 + dx_3^2$ &  $d{\tilde s}^2= \frac{1}{l^2 \Gamma} \big[(l^4 + {\tilde x}_3^2) d{\tilde x}_2^2 $ \\
& & $y=x_2 e^{-x_1},$  &  $-2  dx_1 (x_2 dx_2- x_3 dx_3)$ &  $+ l^4 d{\tilde x}_1^2+(l^4 + {\tilde x}_2^2) d{\tilde x}_3^2 $ \\
&  & $z= x_1$ & $+(1+ x_2^2 + x_3^2) dx_1^2\big],$  &  $+2 {\tilde x}_2 {\tilde x}_3  d{\tilde x}_2 d{\tilde x}_3\big], $ \\
&  &               &  $B=0$              & ${\tilde B}=  \frac{1}{\Gamma} \big[{\tilde x}_2  d{\tilde x}_1 \wedge  d{\tilde x}_2 $  \\
&  &              &                     &  $-{\tilde x}_3  d{\tilde x}_1 \wedge  d{\tilde x}_3\big]$  \\\hline
				
	\end{tabular}}}
\end{center}
\vspace{-2mm}
{\scriptsize
    $\Delta=\alpha_0^2 +{\tilde x}_1^2 +{\tilde x}_2^2, ~ W=x_2^2 +x_3^2 + 2 e^{-2 a x_1},~ \Xi =l^4 a^2 +2(a^2+1)({\tilde x}_2^2+{\tilde x}_3^2),~ \Sigma = 2{\tilde x}_1^2+ 4{\tilde x}_2^2 +{\tilde x}_3^2 +2, \\
	\delta_1=l^4 a^2 +2a^2 {\tilde x}_3^2+2{\tilde x}_2^2 -4a {\tilde x}_2 {\tilde x}_3,~~\delta_2=l^4 a^2 +2a^2 {\tilde x}_2^2+2{\tilde x}_3^2 +4a {\tilde x}_2 {\tilde x}_3,~
	\delta_3=(a {\tilde x}_3- {\tilde x}_2) (a {\tilde x}_2+ {\tilde x}_3),\\
    \Gamma=l^4 +{\tilde x}_2^2 +{\tilde x}_3^2.$}
\\

\begin{center}
	\small {{{\bf Table 8.}~ Non-Abelian T-dual spaces of Lorentzian Thurston geometries.}}
	{\scriptsize
		\renewcommand{\arraystretch}{1.5}{
			\begin{tabular}{| l|l|l| l| l|} \hline \hline
				Geometry${ \large \diagup}$  & Constant matrix & Coordinate & Original& Dual \\
				Isometry subgroup & $E_{_0}(e)$ & transformation &$\sigma$-model & $\sigma$-model \\ \hline
				
				\multirow{5}{*}{$M^3 {\Big / } II$}&\multirow{5}{*}{ $\left(\begin{array}{ccc}
						0 & 0 &  1 \\
						0 & 1 &  0 \\
						1 & 0 &  1 \\
					\end{array} \right)$}
&   $x=x_1 -\frac{1}{6} x_3^3, $  &  $ds^2= dx_2^2 + dx_3^2$ &  $d{\tilde s}^2= -(1+ {\tilde x}_1^2) d{\tilde x}_1^2  $ \\
				  & & $y=x_1+x_3 -\frac{1}{6} x_3^3,$  &  $+2  dx_3 (dx_1- x_3 dx_2),$ &  $+d{\tilde x}_2^2 +2 d{\tilde x}_1 d{\tilde x}_3,$ \\
				&  & $z=x_2 -\frac{1}{2} x_3^2$ & $B=0.$  & ${\tilde B}=- {\tilde x}_1~  d{\tilde x}_1  \wedge  d{\tilde x}_2.$    \\
				&  &  &   &     \\
				
				\multirow{5}{*}{$M^3 {\Big / } III$}&\multirow{5}{*}{ $\left(\begin{array}{ccc}
						0 & 2 &  0 \\
						2 & 0 &  0 \\
						0 & 0 &  \lambda_{_0}^2 \\
					\end{array} \right)$}
        &   $x=-x_2 +e^{x_1}, $  &  $ds^2= \lambda_{_0}^2 dx_3^2$ &  $d{\tilde s}^2= \frac{1}{\lambda_{_0}^2} d{\tilde x}_3^2  $ \\
	&   & $y=\lambda_{_0} x_3,$  &  $+4 e^{x_1}  dx_1  dx_2,$ &  $-\frac{4}{{\tilde x}_2^2-4}  d{\tilde x}_1 d{\tilde x}_2,$ \\
				&  & $z=x_2 +e^{x_1}$ & $B=0$  & ${\tilde B}=\frac{{\tilde x}_2}{{\tilde x}_2^2-4}~  d{\tilde x}_1  \wedge  d{\tilde x}_2$    \\
				&  &  &   &     \\
	
				\multirow{5}{*}{$M^3 {\Big / } IV$}&\multirow{5}{*}{ $\left(\begin{array}{ccc}
						1 & 1 &  -1 \\
						1 & 1 &  0 \\
						-1 & 0 &  0\\
					\end{array} \right)$}
&   $x=-\frac{1}{2} e^{x_1} $  &  $ds^2= dx_2^2 -2 dx_1  dx_3$ &  $d{\tilde s}^2= \frac{1}{\Delta_3}\big[({\tilde x}_2 +{\tilde x}_3)^2 d{\tilde x}_3^2$ \\
&   & $-\frac{e^{-x_1}}{2}(2x_3+x_2^2),$  &  $+(1+x_2^2 +2x_3) dx_1^2$ &  $$-$2\big(1+{\tilde x}_2{\tilde x}_3+{\tilde x}_3^2\big) d{\tilde x}_2 d{\tilde x}_3$ \\
		&  & $y=x_1+x_2,$ &    $-2(x_2 -1)dx_1 dx_2,$    & $+2 d{\tilde x}_1 d{\tilde x}_3\big] +d{\tilde x}_2^2,$    \\
		&  & $z=\frac{1}{2} e^{x_1}$ &  $B=0$  &   ${\tilde B}=\frac{1}{\Delta_3}\big[{\tilde x}_3~  d{\tilde x}_1  \wedge  d{\tilde x}_3$  \\	
		&  & $-\frac{e^{-x_1}}{2}(2x_3+x_2^2)$ &   &  $-({\tilde x}_2+2 {\tilde x}_3)~  d{\tilde x}_2  \wedge  d{\tilde x}_3\big]$   \\		
&  &  &   &     \\

				\multirow{5}{*}{$M^3 {\Big / } V$}&\multirow{5}{*}{ $\left(\begin{array}{ccc}
						0 & 0 &  -1 \\
						0 & 1 &  0 \\
						-1 & 0 &  0\\
					\end{array} \right)$}
&   $x=-\frac{1}{2} e^{x_1} $  &  $ds^2= (2x_3+x_2^2) dx_1^2$ &  $d{\tilde s}^2= d{\tilde x}_2^2 +\frac{1}{\Delta_3}\big[{\tilde x}_2^2 d{\tilde x}_3^2$ \\
&   & $-\frac{e^{-x_1}}{2}(2x_3+x_2^2),$  &  $-2 dx_1(x_2 dx_2 +dx_3)$ &  $~~~~-2 {\tilde x}_2 {\tilde x}_3~ d{\tilde x}_2 d{\tilde x}_3$ \\
		&  & $y=-\frac{1}{2} e^{x_1}$ &    $+ dx_2^2,$    & $~~~~~+2 d{\tilde x}_1 d{\tilde x}_3\big],$    \\
		&  & $+\frac{e^{-x_1}}{2}(2x_3+x_2^2),$ &  $B=0$  &   ${\tilde B}=\frac{1}{\Delta_3}\big[{\tilde x}_3~  d{\tilde x}_1  \wedge  d{\tilde x}_3$  \\	
		&  & $z=x_2$ &   &  $-{\tilde x}_2~  d{\tilde x}_2  \wedge  d{\tilde x}_3\big]$   \\		
&  &  &   &     \\

				\multirow{4}{*}{$M^3 { \large \diagup} VI_0$  }&\multirow{4}{*}{$\left(\begin{array}{ccc}
						-1 & 0 &  0 \\
						0 & 1 &  0 \\
						0 & 0 & \gamma^2\\
					\end{array} \right)$}
&   $x=-{x_1}, $  &  $ds^2= -dx_1^2 + dx_2^2$ &  $d{\tilde s}^2= \frac{1}{\Delta_4} \big[($-$\gamma^2 $-$ {\tilde x}_1^2) d{\tilde x}_1^2 $ \\
& & $y=x_2,$  &  $+ \gamma^2  dx_3^2,$ &  $+ (\gamma^2 - {\tilde x}_2^2) d{\tilde x}_2^2 +  d{\tilde x}_3^2 $ \\
&  & $z= 1+ \gamma x_3$ &   &  $+2 {\tilde x}_1 {\tilde x}_2  d{\tilde x}_1 d{\tilde x}_2\big], $ \\
&  &               &  $B=0$              & ${\tilde B}=  \frac{1}{\Delta_4} \big[-{\tilde x}_2  d{\tilde x}_1 \wedge  d{\tilde x}_3 $  \\
&  &              &                     &  $+{\tilde x}_1  d{\tilde x}_2 \wedge  d{\tilde x}_3\big]$  \\

				\multirow{4}{*}{$M^3 { \large \diagup} VII_0$  }&\multirow{4}{*}{$\left(\begin{array}{ccc}
						1 & 0 &  0 \\
						0 & 1 &  0 \\
						0 & 0 & \sigma^2(1-\lambda^2)\\
					\end{array} \right)$}
&   $x=\sqrt{\lambda^2-1} $  &  $ds^2= dx_1^2 + dx_2^2$ &  $d{\tilde s}^2= \frac{-1}{\Delta_5} \big[2 {\tilde x}_1 {\tilde x}_2  d{\tilde x}_1 d{\tilde x}_2 $ \\
& & $\times (\lambda \sigma x_3-{x_1}),$  &  $+ \sigma^2(1-\lambda^2)  dx_3^2,$ &  $-(\sigma^2(\lambda^2-1) - {\tilde x}_1^2) d{\tilde x}_1^2 $ \\
&  & $y=\lambda x_1$ &   &  $-(\sigma^2(\lambda^2-1) - {\tilde x}_2^2) d{\tilde x}_2^2$ \\
&  &   $+\sigma (1-\lambda^2) x_3,$         &             & $+ d{\tilde x}_3^2 \big], $  \\
&  &   $z=-\lambda\sigma +x_2$           &        $B=0$               &  ${\tilde B}=  \frac{1}{\Delta_5} \big[{\tilde x}_2  d{\tilde x}_1 \wedge  d{\tilde x}_3 $  \\
&  &              &                       &  $-{\tilde x}_1  d{\tilde x}_2 \wedge  d{\tilde x}_3\big] $  \\\hline

				\multirow{4}{*}{Nil $ { \large \diagup} II$  }&\multirow{4}{*}{$\left(\begin{array}{ccc}
						-\frac{l^2}{4} & 0 &  0 \\
						0 & \frac{l^2}{4} &  0 \\
						0 & 0 & \frac{l^2}{4}\\
					\end{array} \right)$}
&   $x=-{x_2}, $  &  $ds^2= \frac{l^2}{4}\big[-dx_1^2 $ &  $d{\tilde s}^2= \frac{l^2}{4(\frac{l^4}{16} +{\tilde x}_1^2)} \big[ d{\tilde x}_2^2 $ \\
& & $y=x_3,$  &  $+(1-x_3^2) dx_2^2 +  dx_3^2$ &  $+  d{\tilde x}_3^2\big]- \frac{4}{l^2} d{\tilde x}_1^2,$ \\
&  & $z= x_1-x_2 x_3$ &  $+2 x_3~dx_1 dx_2\big],$ &   \\
&  &               &  $B=0$              & ${\tilde B}=  \frac{{\tilde x}_1}{(\frac{l^4}{16} +{\tilde x}_1^2)}  d{\tilde x}_2 \wedge  d{\tilde x}_3 $  \\\hline

	\end{tabular}}}
\end{center}
\vspace{6mm}

\begin{center}
	\small {{{\bf Table 8.}~ Continued.}}
	{\scriptsize
		\renewcommand{\arraystretch}{1.5}{
			\begin{tabular}{| l|l|l| l| l|} \hline \hline
				Geometry${ \large \diagup}$  & Constant matrix & Coordinate & Original& Dual \\
				Isometry subgroup & $E_{_0}(e)$ & transformation &$\sigma$-model & $\sigma$-model \\ \hline
				
				\multirow{4}{*}{New Sol $ { \large \diagup} VI_0$  }&\multirow{4}{*}{$\left(\begin{array}{ccc}
						-l^2& 0 &  0 \\
						0 & l^2 &  0 \\
						0 & 0 & l^2 \\
					\end{array} \right)$}
&   $x={x_3} e^{-x_1}, $  &  $ds^2={l^2}\big[dx_2^2 +  dx_3^2 $ &  $d{\tilde s}^2= \frac{l^2}{\Delta_6} \big[d{\tilde x}_1^2- d{\tilde x}_2^2 $ \\
& & $y={x_2} e^{x_1},$  &  $+(x_2^2 +x_3^2-1)dx_1^2$ &  $ +\frac{1}{l^4}({\tilde x}_3 d{\tilde x}_2 +{\tilde x}_2 d{\tilde x}_3)^2$ \\
&  & $z= -x_1$ &  $+2 dx_1 (x_2 dx_2-x_3 dx_3)\big],$ & $- d{\tilde x}_3^2\big],$  \\
&  &               &  $B=0$              & ${\tilde B}=  \frac{1}{\Delta_6}  \big[{\tilde x}_3~d{\tilde x}_1 \wedge  d{\tilde x}_3$  \\
&  &               &                & $-{\tilde x}_2~d{\tilde x}_1 \wedge  d{\tilde x}_2\big]$  \\\hline

				\multirow{4}{*}{Lorentz Sol $ { \large \diagup} VI_0$  }&\multirow{4}{*}{$\left(\begin{array}{ccc}
						l^2& -l^2 &  2l^2 \\
						-l^2 & l^2 &  0 \\
						2l^2 & 0 & 3l^2 \\
					\end{array} \right)$}
&   $x=x_1 +x_2  $  &  $ds^2=l^2\big[e^{2x_3} (dx_1^2 + dx_2^2)$ &  $d{\tilde s}^2= \frac{1}{l^2 \Delta_1} \big[(l^4-{\tilde x}_2^2) d{\tilde x}_2^2 $ \\
& & $+e^{x_3},$  &  $+3 dx_3^2 -2 e^{2x_3} dx_1~dx_2$ &  $ -(3l^4+\tilde{x}_1^2)d\tilde x_1^2$ \\
&  & $y=-x_1 +x_2$ &  $+4 \cosh x_3 dx_1 dx_3 $ & $-2(3l^4-\tilde{x}_1 \tilde{x}_2) d\tilde x_1 d\tilde x_2$  \\
&  &    $+e^{-x_3},$    &  $-4\sinh x_3 dx_2 dx_3\big],$              & $+4 l^4 d\tilde x_3 (d\tilde x_1+ d\tilde x_2)\big],$  \\
&  &     $z = x_3$       &   $B=0$                & ${\tilde B}= \frac{1}{\Delta_{_1}} \big[$-$2 \tilde x_1d\tilde x_1\wedge d\tilde x_2 $ \\
&  &                     &                    & $- (\tilde{x}_1+\tilde{x}_2)(d\tilde x_1\wedge d\tilde x_3$ \\
&  &                     &                    & $+ d\tilde x_2\wedge d\tilde x_3)\big]$ \\
&  &                     &                    &  \\

				\multirow{4}{*}{Lorentz Sol $ { \large \diagup} VI_3$  }&\multirow{4}{*}{$\left(\begin{array}{ccc}
						0& -2 l^2 &  -2l^2 \\
						-2 l^2 & \frac{l^2}{9} &  -\frac{l^2}{9} \\
						-2l^2 & -\frac{l^2}{9} & \frac{l^2}{9} \\
					\end{array} \right)$}
&   $x=x_2 +x_3$  &  $ds^2=-4 l^2 e^{2x_1} \big[dx_1  dx_2$ &  $d{\tilde s}^2= \frac{9}{4 l^2 \Delta_{_2}} \Big[\delta_{_{+}} \delta_{_{-}} d\tilde x_2^2$ \\
& & $+ \frac{e^{6 x_1}}{6}  (x_2 -x_3)^2,$  &  $-\frac{1}{36} e^{6x_1} (dx_2-dx_3)^2$ &  $ +\lambda_{_{+}} \lambda_{_{-}} d\tilde x_3^2 -2\big[3 (\tilde{x}_2^2 + \tilde{x}_3^2)$ \\
&  & $y=$-$\frac{e^{6 x_1}}{3}  (x_2-x_3),$ &  $+ dx_1 dx_3\big], $ & $+4 l^4-10\tilde{x}_2\tilde{x}_3\big]  d\tilde x_2  d\tilde x_3\Big] $  \\
&  &    $z = -2 x_1$    &                & $-\frac{2 l^2}{\Delta_{_2}}\big(d\tilde x_1 d\tilde x_2 +d\tilde x_1 d\tilde x_3\big),$  \\
&  &           &   $B=0$                & ${\tilde B}= \frac{18(\tilde{x}_2 - \tilde{x}_3)}{\Delta_{_2}}~d\tilde x_2\wedge d\tilde x_3$ \\
&  &                     &                    & $-\frac{(\tilde{x}_2 + \tilde{x}_3)}{2 \Delta_{_2}}\big[d\tilde x_1\wedge d\tilde x_2 $ \\
&  &                     &                    & $+d\tilde x_1\wedge d\tilde x_3\big]$ \\\hline

				\multirow{4}{*}{Third Sol $ { \large \diagup} VI_0$  }&\multirow{4}{*}{$\left(\begin{array}{ccc}
						l^2& 0 & 0 \\
						0 & 0 &  -l^2 \\
						0 & -l^2 &-l^2 \\
					\end{array} \right)$}
&   $x=-x_2 e^{x_1}, $  &  $ds^2=l^2\big[-dx_3^2$ &  $d{\tilde s}^2= \frac{l^2}{\Delta_7} \big[d{\tilde x}_1^2+ d{\tilde x}_2^2 $ \\
& & $y=-x_3 e^{-x_1},$  &  $+(1+2 x_2 x_3 -x_3^2)dx_1^2$ &  $ -\frac{1}{l^4} (\tilde{x}_3 d\tilde x_2+\tilde{x}_2 d\tilde x_3)^2 $ \\
&  & $z=x_1$ &         $+2(x_3 -x_2)dx_1 dx_3$ & $-2 d\tilde{x}_2 d\tilde x_3\big]$  \\
&  &                &  $+2 dx_2(x_3 dx_1-dx_3)\big],$              &  \\
&  &                &   $B=0$                & ${\tilde B}= \frac{1}{\Delta_{7}} \Big[\tilde x_2 d\tilde x_1\wedge d\tilde x_3$ \\
&  &                     &                    & $- (\tilde{x}_2+\tilde{x}_3) ~d\tilde x_1\wedge d\tilde x_2\Big]$ \\\hline

				\multirow{4}{*}{Heisenberg $ { \large \diagup} II$} &\multirow{4}{*}{$\left(\begin{array}{ccc}
						\frac{l^2}{4} & 0 & 0 \\
						0 & \frac{l^2}{4} &  \frac{\beta l^2}{4} \\
						0 & \frac{\beta l^2}{4} &\frac{(\beta^2-1) l^2}{4} \\
					\end{array} \right)$}
&   $x=x_3 $  &  $ds^2=\frac{l^2}{4}\big[-dx_3^2$ &  $d{\tilde s}^2= \frac{4l^2}{\Delta_8} \big[(1-\beta^2)d{\tilde x}_2^2 $ \\
{Lorentz-  }   & & $y=-\alpha-x_2$  &  $+(dx_2+\beta dx_3)^2$ &  $ -d\tilde x_3^2 +2 \beta d\tilde{x}_2 d\tilde x_3\big] $ \\
&  &             $-\beta x_3,$ &         $+(x_3 dx_2 -dx_1)^2\big],$ & $+\frac{4}{l^2} d\tilde{x}_1^2,$  \\
&  &            $z=-x_1-\frac{\beta}{2} x_3^2$           &   $B=0$               & ${\tilde B}= -\frac{16 \tilde x_1 }{\Delta_{8}} ~ d\tilde x_2 \wedge d\tilde x_3$ \\\hline

	\end{tabular}}}
\end{center}
\vspace{-2mm}
{\scriptsize $\Delta_1=4l^4-(\tilde{x}_1+\tilde{x}_2)^2,~~\Delta_2=4l^4-(\tilde{x}_2 + \tilde{x}_3)^2, ~~ \Delta_3={\tilde x}_3^2 -1,~~
\Delta_4=\gamma^2 +{\tilde x}_1^2 -{\tilde x}_2^2,\\
\Delta_5=\sigma^2(\lambda^2-1) -{\tilde x}_1^2 -{\tilde x}_2^2,
\Delta_6={\tilde x}_2^2 +{\tilde x}_3^2-l^4,~\Delta_7=l^4 + {\tilde x}_2^2 +2 {\tilde x}_2 {\tilde x}_3,~\Delta_8=l^4 - 16{\tilde x}_1^2,
\\
\delta_{_{\pm}}=2l^2 \pm (\tilde{x}_2 -3 \tilde{x}_3),~ \lambda_{_{\pm}}=2l^2 \pm (3\tilde{x}_2 - \tilde{x}_3).$}


\section{\label{Sec.V} On the conformal invariance of the T-dual $\sigma$-models}

First of all, it is important to note that in order to analyze a modified Ricci flow of the Thurston conjecture, it was presented
a string-inspired three-dimensional Euclidean field theory \cite{Gegenberg2} \footnote{In addition to Ref. \cite{Gegenberg2},
there is an extensive literature on the relation between beta-equations of the NS-NS sector and generalized Ricci flow.
The certain recent mathematical advances in the theory of Ricci flows and their relevance for renormalization group flows have been discussed in \cite{Oliynyk}.
There, the RG flow equations are written in terms of beta-functions which are components of a vector field tangent to the flow on the space of coupling constants of the theory. The beta-functions can be computed in a loop expansion such that in two-dimensional bosonic $\sigma$-model, the loop expansion parameter is $\alpha'$, the square of the string scale. Note that cut-off independence of the regulated quantum theory leads to renormalization group flow equations
\cite{{Oliynyk},{G.Papadopoulos}}.
In this regard, we point out that the renormalization group flow of the $E$-matrix in
the doubled formulation of PL T-duality has calculated in \cite{K.Siampos} (see, also, \cite{Klimcik.Affine}).}.
The theory included a dilaton, an anti-symmetric tensor field and a Maxwell-Chern-Simons field, in addition to the metric.
For a constant dilaton, the theory was appeared to obey a Birkhoff theorem which allowed only nine possible classes
of solutions, depending on the signs of the parameters in the action, in such a way that eight of
them corresponded to the eight Thurston geometries.
In Ref. \cite{Gegenberg2}, in search of a single theory from which all eight of the Thurston geometries arise, it has been turned to the low energy
limit of three-dimensional string theory, which corresponding action is given by
\begin{eqnarray}
S&=& \int d^3 x \Big[\sqrt{-G}~ e^{-2 \Phi} \Big(-\chi + {\cal R}+ 4{\nabla}_{_\mu} \Phi {\nabla}^{^\mu} \Phi-\frac{\epsilon_{_H}}{12} H_{\mu \nu \rho} H^{\mu\nu\rho}
-\frac{\epsilon_{_F}}{2} F_{\mu\nu} F^{\mu\nu} \Big)~~~~\nonumber\\
&&~~~~~~~~~~~~~~~~~~~~~~~~~~~~~~~~~~~~~~~~~~~~~~~~~~~~~~~~~~~~~~~~~~~~~~~~~~-\frac{e}{2} \epsilon^{\mu\nu\rho} A_\mu F_{\nu \rho}\Big],\label{5.1}
\end{eqnarray}
where the last term is the Abelian Chern-Simons term for the Abelian gauge field $A_{(1)}$ with field strength $F_{(2)} =d A_{(1)}$.
In action \eqref{5.1},  ${\cal R}$ is the scalar curvature of the metric $G_{_{\mu\nu}}$, $G=\det G_{\mu\nu}$, $H_{_{\mu\nu\rho}}$ is the field strength of the field  $B_{_{\mu\nu}}$,
and $\Phi(x)$ stands for the dilaton field. Additionally, $\chi, \epsilon_{_H}, \epsilon_{_F}$ and $e$ are some constant parameters.
We remark that if one sets $\chi=-2 \Lambda$, $\epsilon_{_H}=1$, $\epsilon_{_F}=e=0$, then, action \eqref{5.1} reduces to the low energy string effective action,
\begin{eqnarray}\label{5.2}
S_{_{eff}} = \int\! d^{^{3}}x \sqrt{-G} ~ e^{-\Phi}  \Big(2 \Lambda+ {\cal R}  +{\nabla}_{_\mu} \Phi {\nabla}^{^\mu} \Phi-\frac{1}{3} H_{\mu \nu \rho} H^{\mu\nu\rho}\Big),
\end{eqnarray}
where $\Lambda$ is the cosmological constant. Note that to go from their conventions to ours one sends $\Phi\rightarrow {\Phi}/{2}$ and $H\rightarrow 2H$.
The equations of motion which follow from this action are \cite{{callan},{A.Sen1},{A.Sen2},{Tseytlin},{c.hull}}
\begin{eqnarray}
{\cal R}_{_{\mu\nu}}-H_{_{\mu\rho\sigma}}  H^{^{\rho\sigma}}_{_{~~\nu}}+{\nabla}_{_\mu} {\nabla}_{_\nu} \Phi&=&0, \label{5.3}\\
{\nabla}^{^\rho} H_{_{\rho\mu\nu}} -  ({\nabla}^{^\rho}\Phi)  H_{_{\mu\nu\rho}}&=& 0,\label{5.4}\\
2 \Lambda + {\nabla}_{_\mu} {\nabla}^{^\mu} \Phi -  {\nabla}_{_\mu} \Phi {\nabla}^{^\mu} \Phi+\frac{2}{3} H_{_{\mu\nu\rho}}
H^{^{\mu\nu\rho}}&=&0,\label{5.5}
\end{eqnarray}
where ${\cal R}_{_{\mu\nu}}$ is the Ricci tensor, and the field strength $H_{_{\mu\nu\rho}}$ is defined as $H_{_{\mu\nu\rho}}=1/2~(\partial_{_{\mu}} B_{_{\nu\rho}}
+\partial_{_{\nu}} B_{_{\rho\mu}}+\partial_{_{\rho}} B_{_{\mu\nu}})$.
In $\sigma$-model language,
the above equations are known as the vanishing of the one-loop beta-function equations. In fact,
the string effective action is connected to the two-dimensional non-linear $\sigma$-model through these equations \cite{callan}.

According to Ref. \cite{Gegenberg2}, all eight Euclidean Thurston geometries are solutions of the equations
of motion of the theory given by action \eqref{5.1} for various values of the parameters $\chi, \epsilon_{_H}, \epsilon_{_F}$ and $e$.
There, it has been shown that the geometries $E^1 \times S^2$, $E^1 \times H^2$, Sol, Nil and $\widetilde{SL}(2 , \mathbb{R})$
satisfy the the equations of motion of the action \eqref{5.1} provided that $\epsilon_{_F} \neq 0$; also $H^3$ is
a solution if $\epsilon_{_H} =-1$. Considering the statement mentioned above, the aforementioned geometries do not satisfy the
field equations \eqref{5.3}-\eqref{5.5}.
We have verified that among Euclidean geometries, only $E^3$ and $S^3$ are only the solutions to the beta-function equations.
\\
$\bullet$~Euclidean space $E^3$ is flat and thus to be a
solution of the beta-function equations up to the one-loop if it requires $B_{_{\mu\nu}}=0$, $\Lambda=0$ and $\Phi=\varphi_{_0}$ for some constant $\varphi_{_0}$.
\\
$\bullet$~
The sphere metric $S^3$ of Table 1 in the presence of the field strength $H_{_{xyz}} =\sin x /2$ with a constant dilaton field
satisfy the one-loop beta-function equations provided that $\Lambda = -1/2$.
\\\\
{\it Conformal invariance conditions of the Lorentzian Thurston geometries}.
We have checked that among the Lorentzian geometries of Table 2, only $M^3$, Lorentz Sol,  Third Sol and $AdS_3$ satisfy the one-loop beta-function equations.
\\
$\bullet$~Lorentzian space $M^3$ is flat and thus to be a
solution of the beta-function equations up to the one-loop if it requires $B_{_{\mu\nu}}=0$, $\Phi=\varphi_{_0}$ and $\Lambda=0$.
\\
$\bullet$~The Lorentz Sol metric is flat in the sense that its scalar curvature vanishes.
One quickly finds that the only non-zero component of Ricci tensor
is ${\cal R}_{_{zz}}=-2$.
Putting these pieces together, one verifies equations \eqref{5.3}-\eqref{5.5} with  $H_{_{xyz}} = 0$, $\Lambda=0$ and
$\Phi =c_{_0} +c_{_1}e^{-z}+2z $ for some constants $c_{_0}$ and $c_{_1}$.
\\
$\bullet$~The Third Sol metric is a three-dimensional pp-wave metric. It is flat in the sense that its scalar curvature vanishes, and
the only non-zero component of Ricci tensor is ${\cal R}_{_{yy}}=2 e^{2 z}$.
The field strength and dilaton filed that make the background conformal are $H_{_{xyz}} = 0$ and $\Phi =c_{_0}+2z$, respectively.
In addition, the cosmological constant is obtained to be non-zero, $\Lambda=2/l^2$.
\\
$\bullet$~ The one-loop conformal invariance of the $AdS_3$ metric was first examined in Ref. \cite{Horowitz2}.
The $AdS_3$ metric represented in Table 2 in the presence of the B-field $B_{xy} = -l^2/z^2$ with
a constant dilaton field satisfy the one-loop beta-function equations provided that $\Lambda=2/l^2$ \cite{ERA2}.
\\\\
{\it Conformal invariance conditions of the dual backgrounds of Thurston geometries}.
With the aim of better understanding the dual backgrounds we shall examine their conformal invariance.
As will be shown below, of all the dual backgrounds shown in Tables 7 and 8, $E^3/VII_0$, Lorentz Sol/$VI_0$ and
all dual backgrounds of $M^3$ satisfy the one-loop beta-function equations. The results including the scalar curvature, the field strength,  the dilaton field
and cosmological constant of the dual backgrounds are summarized in Table 9.

In order to better understand of the dual spacetimes structure of $M^3$ geometry and also some their physical interpretations
we use some coordinate transformations that make the backgrounds simpler.
Of the dual geometries of $M^3$, the cases of $M^3/III$, $M^3/VI_0$ and $M^3/VII_0$ are the most interesting.
\\
$\bullet$ In the case of the dual geometry of $M^3$ with respect to the $III$
we encounter a non-Abelian self-duality. For this purpose, one may use the change of coordinates
\begin{eqnarray}
\tilde x_1=t-x,~~~~~~\tilde x_2=-2\coth(\frac{t+x}{2}),~~~~~~~~~\tilde x_3=\lambda_{_0} y,\label{5.6}
\end{eqnarray}
to obtain the dual background to the $M^3/III$ (Table 8) in the following form
\begin{eqnarray}
ds^2&=&-dt^2+dx^2+dy^2,\nonumber\\
B&=&\coth(\frac{t+x}{2})~dx\wedge dt, \label{5.7}
\end{eqnarray}
for which the corresponding field strength is found to be zero. As can be seen, the resulting metric is nothing but $M^3$. In fact,
the metric is unchanged under non-Abelian T-duality, and thus we have examined an example of non-Abelian self-duality.
\\
\begin{center}
	\small {{{\bf Table 9.}~ Conformal invariance conditions of the dual backgrounds.}}
	{\scriptsize
		\renewcommand{\arraystretch}{1.5}{
			\begin{tabular}{| l|l|l| l| l|} \hline \hline
				Dual background  &  Scalar curvature & Field strength,  & Dilaton field, & Cosmological\\
				symbol & of the dual metric, ${\cal {\tilde R}}$ & ${\tilde H}_{_{xyz}}$ &$\tilde \Phi$ & constant, $\tilde \Lambda$ \\ \hline				
			
\multirow{2}{*}{$\widetilde{E^3{ \large \diagup} VII_0}$}
& \multirow{2}{*} {$\frac{2{\alpha_{_0}}^2 (7 {\alpha_{_0}}^2 -2 \Delta)}{\Delta^2}$}  & \multirow{2}{*}{$\frac{{\alpha_{_0}}^2 }{\Delta^2}$} & \multirow{2}{*}{$ c_{_0}-\log(\Delta)$}  &  \multirow{2}{*} {$0$} \\
& &  &  &  \\\hline

\multirow{2}{*}{$\widetilde{M^3/II}$  }
&\multirow{2}{*}{$0$} & \multirow{2}{*}{$0$} &  \multirow{2}{*}{$ c_{_0}+c_{_1} {\tilde x}_1$}  &  \multirow{2}{*} {$0$} \\
&  &  & &  \\\hline

\multirow{2}{*}{$\widetilde{M^3/III}$  }
&\multirow{2}{*}{$0$} & \multirow{2}{*}{$0$} &  \multirow{2}{*}{$ c_{_0}+c_{_1} \log(\frac{2-{\tilde x}_2}{2+{\tilde x}_2}) $}  &  \multirow{2}{*} {$0$} \\
&  &  & &  \\\hline

\multirow{2}{*}{$\widetilde{M^3/IV}$  }
&\multirow{2}{*}{$0$} & \multirow{2}{*}{$0$} &  \multirow{2}{*}{$ c_{_0} +\frac{c_{_1}}{2} \log(\frac{1-{\tilde x}_3}{1+{\tilde x}_3})+\log(1-{\tilde x}_3^2)$}  &  \multirow{2}{*} {$0$} \\
&  &  & &  \\\hline

\multirow{2}{*}{$\widetilde{M^3/V}$  }
&\multirow{2}{*}{$0$} & \multirow{2}{*}{$0$} &  \multirow{2}{*}{$  c_{_0} +\log(1-{\tilde x}_3^2)$}  &  \multirow{2}{*} {$0$} \\
&  &  & &  \\\hline

\multirow{2}{*}{$\widetilde{M^3/VI_0}$  }
&\multirow{2}{*}{$\frac{2(2\Delta_4-7 \gamma^2)}{\Delta_4^2}$} & \multirow{2}{*}{$\frac{\gamma^2}{\Delta_4^2}$} &  \multirow{2}{*}{$ c_{_0} -\log(\Delta_4)$}  &  \multirow{2}{*} {$0$} \\
&  &  & &  \\\hline

\multirow{2}{*}{$\widetilde{M^3/VII_0}$  }
&\multirow{2}{*}{$\frac{2\big(2\Delta_5-7 \sigma^2(\lambda^2-1)\big)}{\Delta_5^2}$} & \multirow{2}{*}{$\frac{\sigma^2(1-\lambda^2)}{\Delta_5^2}$} &  \multirow{2}{*}{$ c_{_0}-\log(-\Delta_5)$}  &  \multirow{2}{*} {$0$} \\
&  &  & &  \\\hline
			
\multirow{2}{*}{$\widetilde{Lorentz~ Sol/VI_0}$  }
&\multirow{2}{*}{$0$} & \multirow{2}{*}{$0$} &  \multirow{2}{*}{$ c_{_0}$}  &  \multirow{2}{*} {$0$} \\
&  &  & &  \\\hline

	\end{tabular}}}
\end{center}
$\\$
$\bullet$ The dual geometry of $M^3$ with respect to the $VI_0$ is shown in Table 8.
As is evident, the corresponding dual metric has apparent singularities at the regions ${\tilde x}_2^2-{\tilde x}_1^2 = \gamma^2$.
In order to get more insight, let us use the following coordinate transformation
\begin{eqnarray}
\tilde x_1=r \sinh \varphi,~~~~~~\tilde x_2=r \cosh \varphi,~~~~~~~~~\tilde x_3=t,\label{5.8}
\end{eqnarray}
then, the dual background becomes
\begin{eqnarray}
ds^2&=&-\frac{1}{r^2-\gamma^2} dt^2+\frac{r^2 \gamma^2}{r^2-\gamma^2}  d\varphi^2+dr^2,\label{5.9}\\
B&=&\frac{r^2}{r^2-\gamma^2}~ dt \wedge d\varphi. \label{5.10}
\end{eqnarray}
This metric defines a geometry with naked singularities at the regions $r=\pm \gamma$, as it can be verified by computing the
curvature scalar which is
\begin{eqnarray}
{\cal R} =\frac{-2(2r^2 +5\gamma^2)}{(r^2-\gamma^2)^2}.\label{5.11}
\end{eqnarray}
It is also easy to check that the conformal invariance conditions, equations \eqref{5.3}-\eqref{5.5}, are satisfied by the dual metric,
zero cosmological constant and
\begin{eqnarray}
H&=&\frac{r \gamma^2}{(r^2-\gamma^2)^2}~ dr \wedge d\varphi \wedge dt,\label{5.12}\\
\Phi&=&\varphi_{_0}-\log(r^2-\gamma^2). \label{5.13}
\end{eqnarray}
We have thus found a solution of the string background equations, which is not a black hole, but has naked singularities
and is dual to the $M^3$ geometry.
According to the above result, the non-Abelian T-duality transformation
relates a solution with no curvature singularity to a solution with two curvature singularities.
\\
$\bullet$ Consider now the dual geometry of $M^3$ with respect to the $VII_0$, i.e. $\widetilde{M^3/VII_0}$.
Analogously, the dual metric has also apparent singularities at the region ${\tilde x}_1^2+{\tilde x}_2^2 = \epsilon^2$, where
$ \epsilon^2 = \sigma^2(\lambda^2-1)$.
If we introduce the new coordinates $(t, \varphi, r)$  instead of $({\tilde x}_1, {\tilde x}_2, {\tilde x}_3)$  by means of the transformation
\begin{eqnarray}
\tilde x_1=r \cos \varphi,~~~~~~\tilde x_2=r \sin \varphi,~~~~~~~~~\tilde x_3=t,\label{5.14}
\end{eqnarray}
then, the background including the metric and $\tilde B$-field will become, respectively,
\begin{eqnarray}
ds^2&=&\frac{1}{r^2-\epsilon^2} dt^2-\frac{r^2 \epsilon^2}{r^2-\epsilon^2}  d\varphi^2+dr^2,\label{5.15}\\
B&=&\frac{r^2}{r^2-\epsilon^2}~ d\varphi \wedge dt. \label{5.16}
\end{eqnarray}
The components of metric \eqref{5.15} are ill defined at the regions $r=\pm \epsilon$.
We can test whether there are true singularities by calculating the scalar curvature, which is
$
{\cal R} =\frac{-2(2r^2 +5\epsilon^2)}{(r^2-\epsilon^2)^2}.
$
In addition, one can show that the conformal invariance conditions are, in this case, satisfied with $\Lambda=0$ and
$\Phi=\varphi_{_0}-\log(r^2-\epsilon^2)$.
It can be useful to comment on the fact that the metric \eqref{5.15} and $B$-field \eqref{5.16} can be turned into
\eqref{5.9} and \eqref{5.10}, respectively, if one applies
the Wick rotations $t \rightarrow i t$ and $\varphi \rightarrow i \varphi$ on \eqref{5.15} and
\eqref{5.16}.


\section{\label{Sec.VI} Conclusions}

We have examined that for the metrics of Thurston geometries the isometry subgroups exist and the metrics can be dualized by
the PL T-duality transformations,
in such a way that we have determined dual backgrounds to these geometries including metric and $B$-field.

As announced in the Introduction section, in order to construct the dualizable metrics we needed a non-Abelian three-dimensional Lie subalgebra
of the isometry Lie algebra that generated
the isometry subgroup acting freely and transitively on the target manifold where
the metric of Thurston geometries was defined. Accordingly,
we found all non-Abelian three-dimensional Lie subalgebras of the isometry Lie algebras of the metrics.
Our results showed that Euclidean geometries did not include Lie subalgebras of the Bianchi type $IV$ and $VI_a$,
while Lorentzian geometries encompassed all Bianchi algebras except for $VII_a$.
For all geometries except for $E^1 \times S^2$, $S^3$ and $SL(2, \mathbb{R})$  there was at least one isometric subgroup whose action on the target manifold
was free and transitive.
Additionally, we showed that for all geometries (except for Sol, New Sol and Third Sol)
that possess isometry group whose dimension is greater than
the dimension of the manifold may have several duals. More precisely, for the metrics that admitted various isometry subgroups,
we could construct several backgrounds dual to the metrics.
By a suitable choice of the constant matrix of the $\sigma$-model we have reconstructed the metrics of Thurston geometries
as backgrounds of non-linear $\sigma$-models on the three-dimensional real Lie groups
of the Bianchi type.
The most interesting was the $M^3$ case whose isometry Lie algebra encompassed a
large number of three-dimensional Lie subalgebras (all the Bianchi algebras except for $VI_a, VII_a$ and $IX$).
Finally, we have classified non-Abelian duals of the Thurston geometries with respect to three-dimensional subgroups of the Bianchi type algebras.
As a result, we have obtained eight different types of dual models in the three-dimensional curved backgrounds for the Euclidean geometries
and for the Lorentzians, our results led to twelve dual models.
As we have shown, some of geometries such as $E^3$, $S^3$, $M^3$, Lorentz Sol,
Third Sol and $AdS_3$ belong to the class of string backgrounds with metric, $B$-field and dilaton field.
Regarding the conformality of dual models we showed that only the dual geometries of
$E^3$ with respect to the $VII_0$, Lorentz Sol  with respect to the $VI_0$ and
all duals of $M^3$ are conformally invariant up to the one-loop order.
For the dual background of $M^3/III$, using a suitable coordinate transformation we showed that the model is the non-Abelian self-dual.
In addition, by calculating the true singularities of the dual backgrounds of $M^3$ with respect to the $VI_0$ and $VII_0$
we concluded that the non-Abelian T-duality transformation has been related a solution with no curvature singularity to a solution with two curvature singularities.

Our results spanned an interesting spectrum of PL T-dual $\sigma$-models described by semi-Abelian Drinfeld doubles
generated by the three-dimensional Lie algebras of the Bianchi type.
These are all non-trivial and interesting examples of PL T-dual models which
help in the intent of providing a general classification of three-dimensional geometries
which in some cases can describe supergravity backgrounds.
Another possible direction of further research is to examine the non-Abelian T-dualization of
physically interesting metrics such as the Minkowski metric $M^5$, the $AdS_4$ and $AdS_5$ spaces.
We intend to address these problems in the future.

\subsection*{Declaration of competing interest}

The authors declare that they have no known competing financial
interests or personal relationships that could have appeared to influence the work reported in this paper.

\subsection*{Acknowledgements}

This work has been supported by the research vice chancellor of Azarbaijan Shahid Madani University under research fund No. 1401/537.


\subsection*{Data availability statement}

No data was used for the research described in the article.
\appendix

\section{Isomorphism transformation between six-dimensional Drinfeld doubles with
the isometry Lie algebras of the $E^3$, $H^3$ and $M^3$ geometries}

In this appendix we give isomorphism transformation between six-dimensional Drinfeld doubles with
the isometry Lie algebras of the $E^3$, $H^3$ and $M^3$ geometries that
are isomorphic as Lie algebras.
Note that the isomorphism transformations we find are not unique; they contain several free parameters.
One can give them in a simple form setting the parameters zero or one.
\\
$\bullet$~{\it {The case of $E^3$ geometry.}}~The isomorphism between the Drinfeld double ${{\cal D} {\cal D}}_5:(IX|I)$ and
the isometry Lie algebra of $E^3$ geometry is given by the following transformation
\begin{eqnarray}\label{A.1}
T_{_1} & = & k_{_4},~~~~~~~~~~~~~~~~~~~~~~  T_{_4} = k_{_3}, \nonumber\\
T_{_2} & = &  k_{_2},~~~~~~~~~~~~~~~~~~~~~~T_{_5} = - k_{_5},\nonumber\\
T_{_3} & = & - k_{_1},~~~~~~~~~~~ ~~~~~~~~~       T_{_6} = - k_{_6},
\end{eqnarray}
where $(T_{_1},\cdots, T_{_6})$ and $(k_{_1},\cdots, k_{_6})$ are generators of the Drinfeld double ${{\cal D} {\cal D}}_5:(IX|I)$ and
the isometry Lie algebra of the $E^3$, respectively.\\\\
$\bullet$~{\it {The case of $H^3$ geometry.}}
The isomorphism between the Drinfeld double ${{\cal D} {\cal D}}_1:(IX|V|b),b>0$ and
the isometry Lie algebra of $H^3$ geometry is given by the following transformation
\begin{eqnarray}\label{A.1}
T_{_1} & = &  k_{_4} + \frac{1}{b} k_{_6},~~~~~~~~~~~~~~~~~~~ T_{_4} = {b} k_{_3} - k_{_5}, \nonumber\\
T_{_2} & = & b  k_{_1}-k_{_3}+ \frac{1}{b} k_{_5},~~~~~~~~~~~T_{_5} = k_{_6},\nonumber\\
T_{_3} & = & b  k_{_2}- k_{_4},~~~~~~~~~~~~~~~~~~~~T_{_6} = - k_{_5},
\end{eqnarray}
where $(T_{_1},\cdots, T_{_6})$ and $(k_{_1},\cdots, k_{_6})$ are generators of the Drinfeld double ${{\cal D} {\cal D}}_1:(IX|V|b),b>0$
and the isometry Lie algebra of the $H^3$, respectively.\\\\
$\bullet$~{\it {The case of $M^3$ geometry.}}~
The isomorphism between the Drinfeld double ${{\cal D} {\cal D}}_7:{(VII_0|IV|b)}_{_{b=1}}$ and
the isometry Lie algebra of $M^3$ geometry is given by the following transformation
\begin{eqnarray}\label{A.1}
T_{_1} & = &  -k_{_5},~~~~~~~~~~~~~~~~~~~~~~~~~~~ T_{_4} = k_{_2} - k_{_5}, \nonumber\\
T_{_2} & = & -  k_{_6},~~~~~~~~~~~~~~~~~~~~~~~~~~~T_{_5} =-k_{_1}-2 k_{_3}+k_{_4} -2 k_{_6},\nonumber\\
T_{_3} & = & - k_{_3}+ k_{_4}- k_{_6},~~~~~~~~~~~~~~T_{_6} = k_{_3} +k_{_6},
\end{eqnarray}
where $(T_{_1},\cdots, T_{_6})$ and $(k_{_1},\cdots, k_{_6})$ are generators of the Drinfeld double ${{\cal D} {\cal D}}_7:{(VII_0|IV|b)}_{_{b=1}}$
and the isometry Lie algebra of the $M^3$, respectively.
\\
\\
{\bf ORCID iDs}
\\
Ali Eghbali ~  https://orcid.org/0000-0001-6076-2179
\\
Mahsa Feizi ~ https://orcid.org/0009-0007-3217-6878
\\
Adel Rezaei-Aghdam ~ https://orcid.org/0000-0003-4754-7911


\end{document}